\documentclass[12pt]{article}
\usepackage[utf8]{inputenc}
\usepackage[margin=1in]{geometry}
\usepackage{setspace}
\usepackage{color}
\usepackage{titling}
\usepackage{graphicx}
\usepackage[bf,margin=20pt,format=hang,font={small,stretch=1.1}]{caption}
\usepackage{tikz}
\usepackage{amsmath,amssymb}
\makeatletter\let\over\@@over\makeatother    
\usepackage{hyperref}
\usepackage{simplewick}

\newcommand\TL{\hfil$\displaystyle{##}$}
\newcommand\TR{$\displaystyle{{}##}$\hfil}
\newcommand\TC{\hfil$\displaystyle{##}$\hfil}
\newcommand\TT{\hbox{##}}
\def\seqalign#1#2{\vcenter{\openup1\jot
  \halign{\strut #1\cr #2 \cr}}}
\def\lbldef#1#2{\expandafter\gdef\csname #1\endcsname {#2}}
\newcommand{\eqn}[3][]{\lbldef{#2}{(\ref{#2})}%
\begin{equation} \eqalign{#3} \label{#2} \end{equation}}
\def\eqalign#1{\vcenter{\openup1\jot
    \halign{\strut\span\TL & \span\TR\cr #1 \cr
   }}}
\def\eno#1{(\ref{#1})}

\DeclareMathOperator{\Tr}{Tr}
\DeclareMathOperator{\Norm}{N}

\renewenvironment{abstract}
 {\normalsize
  \begin{center}
   \bfseries \abstractname\vspace{-.5em}\vspace{0pt}
  \end{center}
  \list{}{
   \setlength{\leftmargin}{0in}%
   \setlength{\rightmargin}{\leftmargin}%
  }%
  \item\relax}
 {\endlist}

\title{Non-local non-linear sigma models}

\author{Steven S.~Gubser,${}^a$ Christian B.~Jepsen,${}^a$ Ziming Ji,${}^a$ Brian Trundy,${}^a$\\ and Amos Yarom${}^{a,b}$}
\date{}

\begin{document}
\begin{titlingpage}

\setlength{\droptitle}{-70pt}
\maketitle
\begin{abstract}
We study non-local non-linear sigma models in arbitrary dimension, focusing on the scale invariant limit in which the scalar fields naturally have scaling dimension zero, so that the free propagator is logarithmic.  The classical action is a bi-local integral of the square of the arc length between points on the target manifold.  One-loop divergences can be canceled by introducing an additional bi-local term in the action, proportional to the target space laplacian of the square of the arc length.  The metric renormalization that one encounters in the two-derivative non-linear sigma model is absent in the non-local case.  In our analysis, the target space manifold is assumed to be smooth and Archimedean; however, the base space may be either Archimedean or ultrametric.  We comment on the relation to higher derivative non-linear sigma models and speculate on a possible application to the dynamics of M2-branes.
\end{abstract}
\vfill
June 2019
\end{titlingpage}

\tableofcontents

\section{Introduction}
\label{INTRODUCTION}

Scalar field theories over the reals with bi-local kinetic terms were introduced in \cite{Fisher:1972zz}, and the recent work \cite{Paulos:2015jfa} provides a useful point of entry into the extensive literature.  Similar field theories over the $p$-adic numbers were considered in \cite{Lerner:1989ty} as a continuum description of Dyson's hierarchical model \cite{Dyson:1968up}.  A unifying point of view on the bi-local $O(N)$ vector model was provided in \cite{Gubser:2017vgc}, showing that the standard large $N$ development can be framed in terms that are largely independent of whether the theory is formulated over the reals or the $p$-adics.  The present work extends the study of bi-local theories to bi-local non-linear sigma models, starting with the action
 \eqn{NLNLsM}{
  S = {\mu^{n-s} \over 2\gamma} \int_{V \times V} 
    {d^n x d^n y \over |x-y|^{n+s}} d(\phi(x),\phi(y))^2 \,,
 }
where $|x-y|$ is the distance function on the $n$-dimensional base space $V$ and $d(\phi(x),\phi(y))$ is the distance function on the target manifold.  In the limit $s \to n$, where the theory \eno{NLNLsM} becomes classically scale invariant, we find logarithmic divergences in one-loop diagrams which can be canceled by counterterms that can be expressed in terms of the target space laplacian of the square of the distance function, together with field redefinitions.\footnote{An exception, as we will see, is when $s$ is an even integer and the base space $V = \mathbb{R}^n$.  Through a procedure we will outline in section~\ref{LOCAL}, one recovers in this case a local non-linear sigma model, and at least for $s=2$ we can use our results to check the standard analysis \cite{Friedan:1980jm} of the one-loop beta function.}

Ricci flatness suppresses the one-loop divergences that we encounter, so in a sense (and with significant caveats) we may claim that we are deriving the vacuum Einstein equations from conformal invariance, as in \cite{Friedan:1980jm}.  Our work was partly motivated by the more recent results of \cite{Huang:2019nog}, which were derived for a nearest neighbor arc length model on the Bruhat-Tits tree---in other words, on the other side of the $p$-adic AdS/CFT duality \cite{Gubser:2016guj,Heydeman:2016ldy} from our results for field theories over the $p$-adic numbers.  However, the particular structure of counterterm we find suggests that renormalization of our theories have less to do with renormalization of the local metric as normally understood (i.e.~Ricci flow) than with an augmentation of the action \eno{NLNLsM} to include the target space laplacian of $d(\phi(x),\phi(y))^2$. 

A conservative expectation is that once non-local terms are allowed in a field theory, they proliferate and the theory becomes non-renormalizable.  Theories with purely quadratic bi-local kinetic terms, as studied in \cite{Fisher:1972zz,Lerner:1989ty} (as well as many subsequent works) avoid such problems through a non-renormalization theorem: If we write
 \eqn{Skin}{
  S = {1 \over 2} \int_V d^n k \, \hat\phi(-k) |k|^s \hat\phi(k) + \int_V d^n x \, U(\phi(x)) \,,
 }
then the claim is that the quadratic bi-local term is never renormalized (at least perturbatively), though the purely local term $U(\phi(x))$ certainly is---and depending on details, derivative terms might be radiatively generated.  Non-local interaction terms vitiate this non-renormalization theorem, and one's suspicions could be renewed that there is no sensible theory.  We will not be able in this work entirely to allay such concerns, because we do not give a demonstration parallel to the one in \cite{Friedan:1980jm} that Ward identities based on diffeomorphism invariance guarantee that loop divergences can only modify the original form of the action.  Indeed, the counterterms we generate at one loop do modify the bi-local action in an unexpected way, but one which appears to be controlled in a derivative expansion, so that higher derivative terms can be radiatively generated at each new order without spoiling results from lower orders.  We will revisit the question of renormalizability in section~\ref{OUTLOOK}.

The organization of the rest of this paper is as follows.  In section~\ref{FOURIER} we present the main results in Fourier analysis that we need, both over the reals and the $p$-adics.  In section~\ref{BILOCAL} we explain how double integrals such as the one in \eno{NLNLsM} can be regulated if divergences arise as $|x-y| \to 0$.  In section~\ref{MODEL} we introduce the classical action for the bi-local non-linear sigma model.  In section~\ref{DIVERGENCES} we discuss loop divergences in general terms, including an introductory account of the non-renormalization property of the kinetic term in \eno{Skin}.  In sections~\ref{TWO}-\ref{FOUR} we investigate the simplest one-loop divergences of the bi-local non-linear sigma model, and then in section~\ref{LAPLACIAN} we argue that all these divergences can be canceled by a laplacian counterterm in place of renormalization of the local metric, together with field redefinitions.  As a byproduct of our analysis, we recover in section~\ref{LOCAL} the usual beta function for the two-derivative theory in two dimensions.  We conclude in section~\ref{OUTLOOK} with a summary of possible future directions.

\section{Fourier transforms}
\label{FOURIER}

In loop calculations we will often need to go back and forth between momentum space expressions similar to the ones presented in \eqref{Skin} and their real space counterparts, using the Fourier transforms
\eqn{VFourier}{
  \phi(x) = \int_V d^n k \, e^{2\pi i k \cdot x} \hat\phi(k) \qquad\qquad
  \hat\phi(k) = \int_V d^n x \, e^{-2\pi i k \cdot x} \phi(x) \,.
 }
The relevant results are fairly similar between real and $p$-adic cases, so we present them together.  When $V = \mathbb{R}^n$, the definitions \eno{VFourier} are entirely standard, and $k \cdot x$ can be understood as the ordinary dot product.  Likewise, in this case, $|x|$ is understood as the standard $L^2$ norm on $\mathbb{R}^n$.  We refer to the real case as Archimedean because the norm $|\cdot|$ has the property that if $0<|x|<|y|$, then there is some $n \in \mathbb{Z}$ such that $|y|<|nx|$.

The simplest $n$-dimensional $p$-adic construction is based on letting $V = \mathbb{Q}_{p^n}$ be the (unique) unramified $n$-dimensional extension of $\mathbb{Q}_p$.  Let $\Norm$ and $\Tr$ be the field norm and field trace with respect to the extension $\mathbb{Q}_{p^n}/\mathbb{Q}_p$.  Then we define $|x| = |\Norm(x)|_p^{1/n}$ where $|\cdot|_p$ is the usual $p$-adic norm.  We will refer to the $p$-adic case as ultrametric because the norm $|\cdot|$ just defined has the property $|x+y| \leq \max\{|x|,|y|\}$.  Next we define $k \cdot x = {1 \over n} \Tr(kx)$.  Note that $k \cdot x \in \mathbb{Q}_p$, so to give meaning to $e^{2\pi i k \cdot x}$ we now only need to define $e^{2\pi i \xi}$ for $\xi \in \mathbb{Q}_p$.  To this end we find the unique $p$-adic integer $\lfloor\xi\rfloor$ such that $\xi-\lfloor\xi\rfloor \in [0,1) \cap \mathbb{Q}$, and we understand that by $e^{2\pi i \xi}$ we really mean $e^{2\pi i (\xi-\lfloor\xi\rfloor)}$.

We are particularly interested in the Fourier transform of powers of $|k|$:
 \eqn{PowerFourier}{
  \int_V d^n k \, e^{2\pi i k \cdot x} |k|^s = {\Gamma_V(n+s) \over |x|^{n+s}} + 
    \text{(contact terms)} \,.
 }
Here $\Gamma_V(s)$ is a meromorphic function of $s$ which can be evaluated as
 \eqn{GammaDef}{
  \Gamma_V(s) = {\zeta_v(s) \over \zeta_v(n-s)}
 }
where we set $v=\infty$ in the Archimedean case and $v=p$ in the ultrametric case, with
 \eqn{ZetaDefs}{
  \zeta_\infty(s) \equiv \pi^{-s/2} \Gamma_\text{Euler}(s/2) \qquad\qquad
  \zeta_p(s) \equiv {1 \over 1-p^{-s}} \,.
 }
Intuitively, $\Gamma_V$ is a variant of the Euler gamma, specific to the choice of $V$, and constructed so as to be the coefficient of the $1/|x|^{n+s}$ term in \eno{PowerFourier}.  In the remainder of our discussion, integrals are over $V$ unless otherwise indicated.

The contact terms in \eno{PowerFourier} are somewhat delicate and dependent on detail.  When $-n < s < 0$, the integral in \eno{PowerFourier} is convergent, and no contact terms are needed.  One can easily check that $\Gamma_V(n+s) \to 0$ as $s \to 0$, so when $s=0$ the power law term goes away and we recover the obvious result
 \eqn{FourierDelta}{
  \int d^n k \, e^{2\pi i k \cdot x} = \delta^n(x) \,.
 }
For $s>0$, the integral in \eno{PowerFourier} diverges, and we need a more careful approach.  A good first step is to understand \eno{PowerFourier} in terms of its action on a test function $\phi\colon V \to \mathbb{R}$:
 \eqn{DsDef}{
  \int d^n k \, e^{2\pi i k \cdot x} |k|^s \hat\phi(k) = D^s \phi(x) \,,
 }
where $D^s$ is some linear map on functions $\phi(x)$.  A suitable class of test functions are so-called Schwartz-Bruhat functions.  When $\phi\colon \mathbb{Q}_{p^n} \to \mathbb{R}$, we require that $\phi$ is locally constant with compact support.   For example, the characteristic function of the $p$-adic integers is a Schwartz-Bruhat function on $\mathbb{Q}_p$.  When $\phi\colon \mathbb{R}^n \to \mathbb{R}$, the test functions are more appropriately called Schwartz functions, and their defining property is that they go to $0$ faster than any power of $|x|$, as do all their derivatives.  An example is a Gaussian.  Both in the real and ultrametric cases, the Fourier transform $\hat\phi(k)$ of a Schwartz-Bruhat function is again a Schwartz-Bruhat function.

With \eno{DsDef} taken as the definition of $D^s$, our task is to find a representation of $D^s$ entirely in position space.  In the ultrametric case for arbitrarily positive $s$, one finds
 \eqn{VladimirovDef}{
  D^s \phi(x) = \Gamma_V(n+s) \int d^n y {\phi(y) - \phi(x) \over |x-y|^{n+s}} \,.
 }
This is the Vladimirov derivative.  In the Archimedean case, the same expression \eno{VladimirovDef} is valid for $0<s<2$.  There is one more easy case to dispose of: even positive integer $s$ for Archimedean $V$.  Then $\Gamma_V(n+s) = 0$, which makes sense in \eno{PowerFourier} because the right hand should be purely distributional, on account of $|k|^s = (k^2)^{s/2}$ being analytic in $k^2$.  Explicitly,
 \eqn{DsEven}{
  D^s \phi(x) = {1 \over (2\pi)^s} (-\square)^{s/2} \phi(x) \qquad
   \text{for positive even $s$} \,,
 }
where $\square = \sum_{i=1}^n \partial_{x^i}^2$.

We are left with the task of defining $D^s$ for Archimedean $V$ and for $s>2$ but not an even integer.  Heuristically, the contact terms in \eno{PowerFourier} are a sum of terms of the form $\square^r \delta^n(x)$, where $0 \leq r < s/2$, with divergent coefficients.  To state this more precisely, we write
 \eqn{VladReg}{
  D^s \phi(x) = \Gamma_V(n+s) \int' d^n y {\phi(y) - \phi(x) \over |x-y|^{n+s}} \,,
 }
where a regulated integral
 \eqn{GenReg}{
  \int' d^n y {G(x,y) \over |x-y|^{n+s}}
 }
is rendered finite (if possible) by allowing the subtraction from $G(x,y)$ of a finite sum of smooth functions of either of the following types:
 \begin{enumerate}
  \item[I.] Pure powers: more precisely, any function whose $y$ dependence comes solely through a factor $|x-y|^\alpha$ where $\alpha$ is a real number.  This is meant to include, through the case $\alpha=0$, functions which have no $y$ dependence.
  \item[II.] Higher partial waves: more precisely, any function of the form $Y(\widehat{y-x}) g(|x-y|)$ where $Y(\hat{z})$ is a spherical harmonic on $S^{n-1}$ other than the $s$-wave.
 \end{enumerate}
Type I functions are never integrable, whereas type II functions may or may not be; so at best there is a unique choice of type I functions that will work, whereas many choices of type II functions are possible.  An alternative approach, generalizing the principle value prescription, is to eschew modifications of the integrand and instead carry out $y$ integration in polar coordinates centered around $x$, as follows.  One first performs the angular integrals.  Then the radial integral is restricted to run from $l$ to $L$.  One next allows the subtraction of an arbitrary finite sum of negative powers of $l$ and/or positive powers of $L$, chosen (if possible) so that the limits $l \to 0$ and $L \to \infty$, taken independently, lead to a finite result.  Doing the angular integration first obviates the need for type II functions, while the ultraviolet and infrared cutoffs, $l$ and $L$, obviate the need for type I.\footnote{The alert reader may notice that the alternative approach using cutoffs is not quite equivalent to adjusting $G(x,y)$ by pure powers of $|x-y|$: For example, if $s$ is a positive even integer and $G(x,y) = |x-y|^s$, then we get a logarithmic divergence that would obviously be canceled using an appropriate type I function but cannot be cured using powers of $l$ and/or $L$ after a cutoff integration.  Because we avoid even integer $s$ as well as functions $G(x,y)$ which grow as positive powers of large separation $|x-y|$, we do not need to specify a resolution to this inequivalence.\label{IntSubtle}}

While the subtractions described can in principle cure either ultraviolet (UV) or infrared (IR) divergences, we will be interested only in applications where UV divergences matter: that is, divergences arising when $|x-y| \to 0$ (with $x$ held fixed).  Type II subtractions are relatively innocuous because they follow automatically from performing angular integrations first; therefore we will use the notation $\int d^n y \dots$ to indicate a $y$ integration with type II subtractions which we usually omit to write explicitly.

Although we have stated our integration prescriptions in the abstract, it is easy to see how to apply them to \eno{VladReg} when $\phi$ is a Schwartz function.  Consider the case $2<s<3$, and set $x=0$ for simplicity.  Then \eno{VladReg} becomes
 \eqn{PrimeExample}{
  \int' {d^n y \over |y|^{n+s}} \left[ \phi(y) - \phi(0) \right] 
   = \int {d^n y \over |y|^{n+s}} \left[ \phi(y) - \phi(0) - y_i \partial_i \phi(0) - 
      {1 \over 2} y_{i_1} y_{i_2} \partial_{i_1} \partial_{i_2} \phi(0) \right] \,.
 }
The extra terms in square brackets on the right hand side of \eno{PrimeExample} evidently render the integral convergent near $y=0$ for $2<s<3$.  The term linear in $y$ is clearly a type II function, and the term quadratic in $y$ is a sum of a type II function proportional to $y_{i_1} y_{i_2} - {y^2 \over n} \delta_{i_1i_2}$ (a $d$-wave term) and a type I function proportional to $y^2$.  If $3 \leq s < 4$, then we would need one additional term in the Taylor series expansion of $\phi$ around $y=0$, and this additional term is a type II function.  In summary, for $2<s<4$, and omitting type II subtractions,
 \eqn{PrimeAgain}{
  D^s \phi(0) = \int {d^n y \over |y|^{n+s}} \left[ \phi(y) - \phi(0) - 
     {y^2 \over 2n} \square\phi(0) \right] \,.
 }
Evidently, if $0<s<2$, a simpler subtraction scheme would work, resulting in \eno{PrimeAgain} with the laplacian term omitted, in agreement with \eno{VladimirovDef}.  

For general $s>0$ (other than positive even integers) and Archimedean $V$,
 \eqn{DsGeneral}{
  D^s \phi(0) = \Gamma_V(n+s) \int {d^n y \over |y|^{n+s}} \left[ \phi(y) - 
    \sum_{r=0}^{\lfloor s/2 \rfloor} y^{2r} b_r \square^r \phi(0) \right]
 }
where
 \eqn{brDef}{
  b_r = {\Gamma_\text{Euler}({n \over 2}) \over 
   2^{2r} \Gamma_\text{Euler}(r + {n \over 2}) \Gamma_\text{Euler}(r+1)} \,.
 }
In principle, one may derive \eno{DsGeneral} by subtracting an appropriate number of terms in the Taylor series expansion of $\phi(y)$ and then finding appropriate type II subtractions to bring the result into the form \eno{DsGeneral}. 

A more efficient way to determine the coefficients $b_r$ is to start from \eno{DsGeneral} and Fourier transform:
 \eqn{BesselInt}{
  &\int d^n x \, e^{-2\pi i k \cdot x} D^s \phi(x) 
   \cr &\qquad{}
   = \Gamma_V(n+s) \int d^n x \, e^{-2\pi i k \cdot x} \int {d^n y \over |x-y|^{n+s}} 
    \left[ \phi(y) - \sum_{r=0}^{\lfloor s/2 \rfloor}
     (x-y)^{2r} b_r \square^r \phi(x) \right]
   \cr &\qquad{}
   = {2 \Gamma_V(n+s) \over \zeta_\infty(n-1)} \int_0^\infty {d\tilde{y} \over \tilde{y}^{s+1}}
     \int_0^\pi d\theta \, (\sin\theta)^{n-2} \hat\phi(k) 
     \left[ e^{2\pi i |k|\tilde{y} \cos\theta} - \sum_{r=0}^{\lfloor s/2 \rfloor}  
      b_r (2\pi i)^{2r} k^{2r} \tilde{y}^{2r} \right]
   \cr &\qquad{}
   = \Gamma_V(n+s) (2\pi)^{{n \over 2} + s} |k|^s \hat{\phi}(k) \int_0^\infty d\rho \, 
     \left[ \rho^{-{n \over 2} - s}
     J_{\frac{n}{2} - 1}(\rho) - \sum_{r=0}^{\lfloor s/2 \rfloor} a_r \rho^{2r-s-1} \right]
 }
In the second equality of \eno{BesselInt}, we have partially carried out the $y$ integral in polar coordinates around the point $y=x$, introducing a radial variable $\tilde{y} = |x-y|$.  In the third equality, we have carried out the angular $\theta$ integral and introduced a new radial variable, $\rho = 2\pi |k|\tilde{y}$.  The $\rho$ integral in the last line of \eno{BesselInt} converges, provided $s$ is positive but not an even integer, and provided the coefficients $a_r$ are coefficients in the Taylor series expansion of the Bessel function around $\rho=0$.  These coefficients $a_r$ are well known, and from them one can recover the expression \eno{brDef} for the $b_r$.

\section{Bi-local integrals}
\label{BILOCAL}

We are particularly interested in double integrals of the form
 \eqn{BilocalInt}{
  \int_{V \times V} {d^n x d^n y \over |x-y|^{n+s}} G(x,y)
 }
where $s>0$ and $G(x,y)$ is piecewise constant if $V$ is ultrametric and smooth if $V$ is Archimedean.  Unless otherwise noted, all double integrals over $x$ and $y$ will by taken over all of $V \times V$.  In the ultrametric case, for any $s>0$, following \eno{VladimirovDef} we define
 \eqn{ultraBIF}{
  \int' {d^n x d^n y \over |x-y|^{n+s}} G(x,y) \equiv 
    \int {d^n x d^n y \over |x-y|^{n+s}} \left[ G(x,y) - G(x,x) \right] \,.
 }
In the Archimedean case, we define
 \eqn{BIF}{
  \int' {d^n x d^n y \over |x-y|^{n+s}} G(x,y)
 }
by performing the $y$ integration first and allowing the subtraction of type I and type II functions to $G(x,y)$ in order to achieve a finite result (if possible).  As in the previous section, type II subtractions are deemed relatively inconsequential, so even unprimed integration over $x$ and $y$ means to perform the $y$ integration first, allowing the subtraction of type II functions in order to achieve a finite result (if possible).  Explicitly, for $s$ not a positive even integer,
 \eqn{BilocalCase}{
  \int' {d^n x d^n y \over |x-y|^{n+s}} G(x,y) &= 
    \int {d^n x d^n y \over |x-y|^{n+s}} \Bigg[ G(x,y)
    - \sum_{r=0}^{\lfloor s/2 \rfloor}
       b_r \square_y^r G(x,y)\Big|_{y=x} (y-x)^{2r}  \Bigg] \,,
 }
where the coefficients $b_r$ are as given in \eno{brDef}.  We avoid positive even integer $s$ when $V$ is Archimedean because in this case we expect that our constructions will lead instead to purely local theories; also, precisely in this case, the subtleties pointed out in footnote~\ref{IntSubtle} regarding logarithmic divergences come into play.

Our computational strategy will turn on converting bi-local position space integrals into Fourier space integrals.  Let's start with the simplest example of that calculation, valid for ultrametric $V$ and any $s>0$, and also for Archimedean $V$ and $0<s<2$.  Let $\phi\colon V \to \mathbb{R}$ be a Schwartz-Bruhat function.  Then
 \eqn{KineticTransform}{
  \int {d^n x d^n y \over |x-y|^{n+s}} &\left[ \phi(x) - \phi(y) \right]^2 
   = \int {d^n x d^n y \over |x-y|^{n+s}} \left( \left[ \phi(x) - \phi(y) \right]^2 + 
     \phi(x)^2 - \phi(y)^2 \right)
   \cr &
   = -2 \int d^n x \, \phi(x) \int {d^n y \over |x-y|^{n+s}}
     \left[ \phi(y) - \phi(x) \right]
   \cr &
   = -{2 \over \Gamma_V(n+s)} \int d^n x \, \phi(x) D^s \phi(x)
   = -{2 \over \Gamma_V(n+s)} \int d^n k \, \hat\phi(-k) |k|^s \hat\phi(k) \,.
 }
The first step is actually the trickiest, because it is not clear from the rules of integration set forth following \eno{BIF} that we are allowed to add a function like $\phi(x)^2 - \phi(y)^2$ to the integrand.  To justify this step, we denote $f(x) = \phi(x)^2$, and we argue that
 \eqn{fJustify}{
  \int {d^n x d^n y \over |x-y|^{n+s}} \left[ \phi(y)^2 - \phi(x)^2 \right] = 
   {1 \over \Gamma_V(n+s)} \int d^n x \, D^s f(x) = 0\,.
 }
The second integral in \eno{fJustify} is the $k=0$ component of the Fourier transform of $D^s f(x)$.  But this Fourier transform is $|k|^s \hat{f}(k)$, and since $s>0$ the $k=0$ component indeed vanishes.

Let's now pursue the same computation for the Archimedean case with $2<s<4$.  On one hand, using \eno{BilocalCase},
 \eqn{RegKT}{
  &\int' {d^n x d^n y \over |x-y|^{n+s}} \left[ \phi(x) - \phi(y) \right]^2 
   = \int {d^n x d^n y \over |x-y|^{n+s}} 
      \left( \left[ \phi(x) - \phi(y) \right]^2 -
       {(y-x)^2 \over n} (\partial\phi(x))^2 \right)
     \cr &\qquad\qquad\quad{}
   = \int {d^n x d^n y \over |x-y|^{n+s}} \left( 
     \left[ \phi(x) - \phi(y) \right]^2 - 
       {(y-x)^2 \over 2n} \left[ -2\phi(x) \square \phi(x) + \square \phi(x)^2 \right]
       \right) \,.
 }
On the other hand, using \eno{PrimeAgain},
 \eqn{PhiReg}{
  -{2 \over \Gamma_V(n+s)} \int d^n k & \, \hat\phi(-k) |k|^s \hat\phi(k)
    = -2\int d^n x \, \phi(x) \int' {d^n z \over |z|^{n+s}} 
      \left[ \phi(x+z) - \phi(x) \right]
     \cr &
    = \int {d^n x d^n y \over |x-y|^{n+s}} \left( 
      2\phi(x) \left[ \phi(x) - \phi(y) \right] +
       {(y-x)^2 \over n} \phi(x) \square \phi(x) \right) \,.
 }
In order to conclude
 \eqn{RegConclude}{
  \int' {d^n x d^n y \over |x-y|^{n+s}} \left[ \phi(x) - \phi(y) \right]^2 = 
   -{2 \over \Gamma_V(n+s)} \int d^n k & \, \hat\phi(-k) |k|^s \hat\phi(k) \,,
 }
we must therefore argue that the final integrals in \eno{RegKT} and \eno{PhiReg} agree.  Subtracting \eno{PhiReg} from \eno{RegKT} and simplifying slightly with the definition $f(x) = \phi(x)^2$, we arrive at
 \eqn{HigherVlad}{
  \int {d^n x d^n y \over |x-y|^{n+s}} \left[ f(y) - f(x) - {(y-x)^2 \over 2n} \square f(x)
    \right] = {1 \over \Gamma_V(n+s)} \int d^n x \, D^s f(x) = 0 \,.
 }
The first equality in \eno{HigherVlad} follows from \eno{PrimeAgain}, and the second is by the same argument used following \eno{fJustify}.  To summarize, for Archimedean $V$ and for $2<s<4$,
 \eqn{TwoFourSummary}{
  \int {d^n x d^n y \over |x-y|^{n+s}} & \left( {(y-x)^2 \over n} (\partial\phi(x))^2 - 
    \left[ \phi(x) - \phi(y) \right]^2 \right)
    \cr &
   = {2 \over \Gamma_V(n+s)} \int d^n k \, \phi(-k) |k|^s \phi(k) \,.
 }
$\Gamma_V(n+s) > 0$ for $2<s<4$, and so without the $(\partial\phi)^2$ on the left hand side of \eno{TwoFourSummary} we would have a sign problem.  The equality \eno{RegConclude} can be checked in a similar manner for $s>4$.  A key relation is
\begin{equation}
	\square_y^m \left(\phi(x) - \phi(y)\right)^2 \Big|_{x=y} = -2 \phi(x) \square^m \phi(x) + \square^m \phi(x)^2\,.
\end{equation}

Two take-away lessons are:
 \begin{itemize}
  \item When we write simple $|k|^s$ kinetic terms in momentum space, in position space we are combining non-local position space terms and local terms involving derivatives in a precisely tuned ratio.
  \item There is some freedom in the precise structure of the position space form, as exemplified by the equality of the last integrals in \eno{RegKT} and \eno{PhiReg} due to a manipulation which is the non-local version of integration by parts.
 \end{itemize}

\section{The bi-local non-linear sigma model}
\label{MODEL}

We are now in a position to present the action for the bi-local non-linear sigma model.  Let $M$ be a smooth $D$-dimensional manifold with a Riemannian metric $g_{ab}$, whose Riemann and Ricci tensors are
 \eqn{CurvatureTensors}{
  R_{ab}{}^c{}_d = \partial_a \Gamma^c_{bd} - \partial_b \Gamma^c_{ad} + 
    \Gamma^c_{ae} \Gamma^e_{bd} - \Gamma^c_{be} \Gamma^e_{ad} \qquad\qquad
  R_{ac} = g^{bd} R_{abcd} \,.
 }
Given any two points $X$ and $Y$ on $M$, let 
 \eqn{Qdef}{
  Q(X,Y) = d(X,Y)^2
 }
be the square of the shortest distance between $X$ and $Y$. Clearly, $Q(X,Y)$ is a smooth function of $X$ and $Y$, provided $X$ and $Y$ are not too far apart.  For smooth functions $\phi\colon V \to M$ whose range is sufficiently localized, we consider the action functional
 \eqn{Sdef}{
  S = {\mu^{n-s} \over 2\gamma} \int' {d^n x d^n y \over |x-y|^{n+s}} Q(\phi(x),\phi(y)) \,,
 }
where $\int'$ indicates a regulated double integral of the type discussed around \eno{ultraBIF}-\eno{BilocalCase}.\footnote{One may wonder whether the primed integral, as defined following \eno{BIF}, spoils coordinate invariance of the integrand. For instance, if $s$ is sufficiently large we may, in light of \eqref{BilocalCase}, be required to include a $\square_y Q$ term to the integrand, which if written only in terms of partial derivatives does not appear to be coordinate invariant.  In fact, it is easy to convince oneself that, e.g., $\square_y Q$ can be constructed from covariant quantities:
$\partial_{y^a} \partial_{y^b} Q = \frac{\partial^2 Q}{\partial \phi^i \partial \phi^j} \frac{\partial \phi^i}{\partial y^a} \frac{\partial \phi^j}{\partial y^b} + \frac{\partial Q}{\partial \phi^i} \frac{\partial^2 \phi^i}{\partial y^a \partial y^b} =\left( \nabla_{\phi^i} \frac{\partial Q}{\partial \phi^j} \right)\frac{\partial \phi^i}{\partial y^a} \frac{\partial \phi^j}{\partial y^b} + \frac{\partial Q}{\partial \phi^i} \frac{\partial \phi^j}{\partial y^a} \nabla_{\phi^j} \frac{\partial \phi^i}{\partial y^b}$.}
Note that this discussion requires us to avoid positive even integer $s$ when $x$ and $y$ are valued in $\mathbb{R}^n$ (as opposed to $\mathbb{Q}_{p^n}$).  In the Archimedean setting, when $s>2$, there are derivative terms like $(\partial\phi)^2$ implicitly built into \eno{Sdef}, with coefficients tuned so as to ensure convergence of the integral.  The parameter $\mu$ has dimensions of energy so that we can regard $\phi$ and $Q(\phi(x),\phi(y))$ as dimensionless. The factor $\gamma$ is a loop-counting parameter: Classical effects are $O(\gamma^{-1})$, one-loop amplitudes are $O(\gamma^0)$, two loop amplitudes are $O(\gamma)$, and so forth.  In other words, $\gamma$ plays the role of $\hbar$.

A close cousin of the action \eno{Sdef} was considered in \cite{Huang:2019nog}:
 \eqn{SNN}{
  S = \sum_{\langle xy \rangle} d(\phi(x),\phi(y))^2
 }
where now $x$ and $y$ are vertices of a graph and the sum is over undirected edges.  The formula actually appears earlier in \cite{Friedan:1980jm}, though it was intended there to be considered on a square lattice, as a regulator for the local non-linear sigma model, rather than on the Bruhat-Tits tree as in \cite{Huang:2019nog}.

We require the range of the maps $\phi$ to be sufficiently localized in order to ensure that we do not encounter any failures of smoothness in $Q(X,Y)$, and in order to ensure that we can use a single system of Riemann normal coordinates for $\phi$ throughout.  One can now solve the geodesic equation perturbatively in the curvature and use that to evaluate $Q(X,Y)$ and expand the action \eqref{Sdef}, 
\eqn{SphiExpand}{
  S = S_2 + S_4 + \dots \,,
 }
where
 \eqn{Sfree}{
  S_2 =  {\mu^{n-s} \over 2\gamma} g_{ab} \int' {d^n x d^n y \over |x-y|^{n+s}}
   \left[ \phi^a(x) - \phi^a(y) \right] \left[ \phi^b(x) - \phi^b(y) \right]
 }
and
 \eqn{Sint}{
  S_4 = -{\mu^{n-s} \over 6\gamma} R_{abcd} \int' {d^n x d^n y \over |x-y|^{n+s}}
    \phi^a(x) \phi^b(y) \phi^c(x) \phi^d(y) \,.
 }
Here $g_{ab}$ and $R_{abcd}$ are evaluated at the origin of the Riemann normal coordinates, which we assume is at $X=0$. Often, the definition of Riemann normal coordinates includes the requirement $g_{ab} = \delta_{ab}$, but this is not necessary for our calculations, and we find it more convenient to retain explicit factors of $g_{ab}$ and the inverse metric $g^{ab}$.  Put differently, we are choosing a coordinate system so that all geodesics passing through the origin are linear in the affine parameter:
\begin{equation}\label{MetricExpand}
	g_{ab}(\phi) = g_{ab} - \frac{1}{3} R_{acbd} \phi^c \phi^d + O(\phi^3)\,.
\end{equation}
The ellipsis in \eno{SphiExpand} indicates higher order interactions, involving five or more powers of $\phi(x)$ and/or $\phi(y)$, as well as derivatives and powers of the curvature.  We will consider up to six point interactions in section~\ref{FOUR}.

\section{Loop divergences in momentum space}
\label{DIVERGENCES}

Our aim in this section is to introduce the main concepts related to divergent loop diagrams that we will need in our analysis of the non-local non-linear sigma model.  As a warmup, we first exhibit the simplest manifestation of the non-renormalization theorem of the non-local quadratic kinetic term in the action \eno{Skin}, where $V = \mathbb{R}^n$ or $\mathbb{Q}_{p^n}$, with
 \eqn{UChoice}{
  U(\phi) = {g \over 3!} \phi^3 \,.
 }
The purely cubic theory is unstable, but it serves our purpose because we are only interested in analyzing the behavior of the one-loop correction to the propagator.  Using the diagram shown in figure~\ref{SimpleDiagrams}a, we obtain the one-loop contribution to the quadratic part of the one-particle irreducible (1PI) effective action:
 \begin{figure}
  \centerline{\begin{tikzpicture}
   \draw[thick] (-0.5,0) -- (0,0);
   \draw[thick] plot [smooth, tension=1] coordinates {(0,0) (0.75,0.6) (1.5,0)};
   \draw[thick] plot [smooth, tension=1] coordinates {(0,0) (0.75,-0.6) (1.5,0)};
   \draw[thick] (1.5,0) -- (2,0);
   \node[below] at (-0.3,0) {$-g$};
   \node[below] at (1.7,0) {$-g$};
   \draw[semithick, ->] (-1,0.3) -- (-0.1,0.3);
   \node[above] at (-0.45,0.3) {$k$};
   \draw[semithick, ->] plot [smooth, tension=1] coordinates {(0.2,0.55) (0.75,0.8) (1.3,0.55)};
   \node[above] at (0.75,0.8) {$\ell$};
   \draw[semithick, ->] plot [smooth, tension=1] coordinates {(0.2,-0.55) (0.75,-0.8) (1.3,-0.55)};
   \node[below] at (0.75,-0.8) {$k-\ell$};
   \node[below=45pt] at (0.75,0) {(a)};
   \draw[thick] (4.2,0) -- (4.8,0);
   \draw[thick] plot [smooth, tension=1] coordinates {(4.5,0) (4.2,0.63) (4.5,1) (4.8,0.63) (4.5,0)};
   \draw[semithick, ->] (4.2,-0.3) -- (4.8,-0.3);
   \node[below] at (4.5,-0.3) {$k$};
   \draw[semithick, ->] plot [smooth, tension=1] coordinates {(4.8,1.15) (4.5,1.23) (4.2,1.15)};
   \node[above] at (4.5,1.23) {$\ell$};
   \node[below=45pt] at (4.5,0) {(b)};
  \end{tikzpicture}}
  \caption{(a) One-loop contribution to the 1PI two-point function.  (b) A diagram where the loop consists of a single propagator closing on itself.}\label{SimpleDiagrams}
 \end{figure}
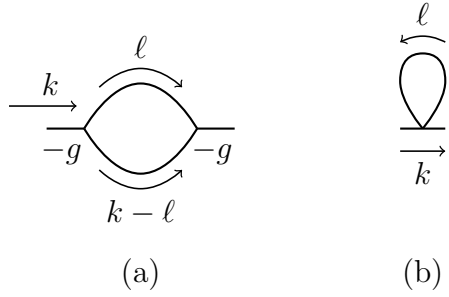
 \eqn{LoopSummary}{
  \delta\Gamma_2(k) = -{g^2 \over 2} I \qquad\text{where}\qquad
  I = \int {d^n\ell \over |\ell|^s |k-\ell|^s} \,.
 }
We continue the convention of integrating over all of $V$ except as otherwise indicated.  Let's assume $n>2s$, so $I$ is UV divergent (and IR convergent).  To regulate the divergence, we introduce a hard cutoff: $|\ell| \leq \Lambda$.  If $V = \mathbb{R}^n$, then $\Lambda$ can be any positive real number.  If $V = \mathbb{Q}_{p^n}$, then we will require that $\Lambda$ is an integer power of $p$.

The ultrametric case is easy to analyze, because when $|\ell| > |k|$ we have $|\ell| = |k-\ell|$ exactly.  So, except in the compact region where $|\ell| \leq |k|$, the integrand has no $k$ dependence at all.  Therefore, any UV divergences are entirely independent of $k$, and to evaluate them we can set $k=0$:
 \eqn{ultraI}{
  I(\Lambda) &= \int_{|\ell|\leq \Lambda} {d^n \ell \over |\ell|^s |k-\ell|^s} = 
    \int_{|\ell|\leq \Lambda} {d^n \ell \over |\ell|^{2s}} + \text{(UV finite)}
    \cr &
   = {\zeta_p(n-2s) \over \zeta_p(n)} \Lambda^{n-2s} + \text{(UV finite)} \,.
 }
The last equality comes from splitting the integration region into shells with fixed $|\ell|$; then the integral becomes a geometric sum.  Because the divergent part of $I(\Lambda)$ has no $k$-dependence, the counterterm required to cancel it is proportional to $\int d^n k \, \hat\phi(-k) \hat\phi(k) = \int d^n x \, \phi(x)^2$.  In other words, it is a mass term.  This argument is easy to generalize to the statement that only purely local terms (powers of $\phi(x)$) can be radiatively generated starting from the action \eno{Skin} over $V = \mathbb{Q}_{p^n}$.  An essentially equivalent argument was made in a Wilsonian picture in \cite{Lerner:1989ty}.

The Archimedean case is more subtle because of the possibility of subleading divergences.  A straightforward approach is to expand
 \eqn{IntExpand}{
  I(\Lambda) = \int_{|\ell|\leq \Lambda} {d^n \ell \over |\ell|^s |k-\ell|^s}
    = \int_{|\ell|\leq \Lambda} {d^n \ell \over |\ell|^{2s}} \left( 1 - 
       {2 k \cdot \hat\ell \over |\ell|} + {k^2 \over \ell^2} \right)^{-s/2}
 }
in powers of $k$.  Terms with an odd number of powers of $k$ vanish by parity, leaving only terms analytic in $k^2$.  Of these, only terms proportional to $k^{2r}$ with $r \leq {n \over 2} - s$ are UV divergent.  In short, the divergent part of $I(\Lambda)$ is a polynomial in $k^2$ whose order is $\left\lfloor {n \over 2} - s \right\rfloor$.  A divergent term proportional to $k^{2r}$ requires a counterterm proportional to $\int d^n x \, \phi(x) \square^r \phi(x)$.  These are the radiatively generated derivative terms mentioned in section~\ref{INTRODUCTION}.

We should note a troublesome feature of the hard momentum cutoff for Archimedean theories: The coefficients one finds for sub-leading divergences depend on how one implements the cutoff.  For example, it is easy to check that the coefficient of the $k^2$ term in $I(\Lambda)$ changes if instead of requiring $|\ell|\leq \Lambda$ we impose the more democratic condition $\left| \ell - {k \over 2} \right| \leq \Lambda$.  However, the feature that we care about, namely the fact that the divergent terms have only polynomial dependence on $k^2$, doesn't depend on the details of the cutoff.  It is perhaps instructive to consider one other alternative, namely dimensional regularization, in which one first computes
 \eqn{Idimen}{
  I = {\Gamma_{\mathbb{R}^n}(2s-n) \over \Gamma_{\mathbb{R}^n}(s)^2} (k^2)^{n-2s \over 2}
 }
by continuing to a domain of $n$ in which the integral is convergent.  (In the current example, $s<n<2s$ is such a domain.)  The only divergences one then tracks are poles of the right hand side of \eno{Idimen} as a function of $n$.  These occur precisely when ${n-2s \over 2}$ is a non-negative integer.  It is characteristic of dimensional regularization that there is (at most) one divergent term for a given $n$, corresponding to a logarithmic divergence in the original integral.

The loop diagrams we will need to consider in our analysis of the non-local non-linear sigma model are simpler than \eno{LoopSummary} in one regard: The loop is a single propagator starting and ending at the same vertex.  This matters because there is then only one internal momentum $\ell$, and imposing the hard cutoff $|\ell| \leq \Lambda$ is a privileged choice because it corresponds to integrating $\ell$ over an $O(n)$-invariant region.  An example is the diagram shown in figure~\ref{SimpleDiagrams}b, which is proportional to
 \eqn{IzeroLoop}{
  I_0 = \int {d^n\ell \over |\ell|^s} \,,
 }
assuming that whatever vertex factor is needed to fully evaluate the diagram doesn't depend on $\ell$.  We also assume $n>s$ so that $I_0$ is UV divergent but IR convergent.  We straightforwardly find
 \eqn{IzeroValue}{
  I_0 = \left\{ \seqalign{\span\TL &\qquad \span\TT}{
    {2 \over \zeta_\infty(n)} {\Lambda^{n-s} \over n-s} & for $V = \mathbb{R}^n$  \cr
    {\zeta_p(n-s) \over \zeta_p(n)} \Lambda^{n-s} & for $V = \mathbb{Q}_{p^n}$\,.}
   \right.
 }
There are obviously no subleading divergences in $I_0$.

For convenience we introduce
 \eqn{EpsilonDef}{
  \epsilon = n-s \,.
 }
We are interested in divergences proportional to $\log\Lambda$ that arise when $\epsilon = n-s = 0$.  As a technical trick to isolate these divergences, we make $\epsilon$ small and positive, and we look for divergences of the form $\Lambda^\epsilon/\epsilon$, which in the $\epsilon \to 0^+$ limit give rise to $\log\Lambda$ terms.  To characterize this limit precisely, given $n_0 > 0$ and $\lambda \in \mathbb{R}$, we set $n = n_0 + \lambda \epsilon$ and $s = n_0 + (\lambda-1) \epsilon$ and then take the $\epsilon \to 0^+$ limit with $n_0$ and $\lambda$ held fixed.  (Clearly then we are allowing non-integer $n$, in the spirit of \cite{Friedan:1980jm}.)  For the most part, our final results are independent of $\lambda$.  When $\epsilon$ is sufficiently small, we may replace \eno{IzeroValue} with
 \eqn{IzeroAgain}{
  I_0 = i_0 {\Lambda^\epsilon \over \epsilon} \qquad\text{where}\qquad
  i_0 = \left\{ \seqalign{\span\TL &\qquad \span\TT}{
    {2 \over \zeta_\infty(n_0)} + O(\epsilon) & for $V = \mathbb{R}^n$  \cr
    {1 \over \zeta_p(n_0) \log p} + O(\epsilon) & for $V = \mathbb{Q}_{p^n}$\,.}
   \right.
 }
If we lift the requirement that $\Lambda$ is an integer power of $p$ when $V = \mathbb{Q}_{p^n}$, then \eno{IzeroAgain} is unaltered, because $I_0 = {\zeta_p(\epsilon) \over \zeta_p(n)} p^{\epsilon \lfloor \log_p \Lambda \rfloor}$, and $p^{\epsilon \lfloor \log_p \Lambda \rfloor}$ differs from $\Lambda$ at most by a factor of $p^\epsilon = 1 + O(\epsilon)$.  The important point is that in the limit $\epsilon \to 0^+$, $I_0$ includes a logarithmic term $i_0 \log \Lambda$, and isolating this term is our stated objective.

We will encounter one other loop integral:
 \eqn{Itwo}{
  I_2(k) = \int d^n \ell \, {|k-\ell|^s \over |\ell|^s} \,.
 }
It comes from graphs similar to the one in \eno{IzeroLoop}, but with a vertex prefactor $|k-\ell|^s$.  Using the same reasoning that led to \eno{ultraI}, we see that when a hard cutoff $|\ell| \leq \Lambda$ is imposed, one obtains in the ultrametric case
 \eqn{ItwoUltra}{
  I_2(k) = \Lambda^n + \text{(UV finite)} \qquad\text{when $V = \mathbb{Q}_{p^n}$} \,.
 }
If $V = \mathbb{R}^n$ and $n$ is positive but not an even integer, then
 \eqn{ItwoDivergences}{
  I_2(k) = \sum_{r=0}^{\lfloor n/2 \rfloor} c_r k^{2r} {\Lambda^{n-2r} \over n-2r} + 
    \text{(UV finite)}
 }
for some coefficients $c_r$.  If we choose $n_0$ positive but not an even integer and fix any finite value of $\lambda$, then for sufficiently small $\epsilon$, \eno{ItwoDivergences} applies, and the least singular power of $\Lambda$ appearing in it is $\Lambda^{n-2\lfloor n/2 \rfloor}$.  As $\epsilon \to 0^+$, this power remains positive and finite, tending to $\Lambda^{n_0-2\lfloor n_0/2 \rfloor}$.  So there is no $\log\Lambda$ behavior, even in the $\epsilon \to 0^+$ limit.  If instead we make $n_0$ a positive even integer, then by choosing the very particular value $\lambda=1$, so that $s=n_0$ exactly, we find (for sufficiently small $\epsilon>0$) that the least singular term in \eno{ItwoDivergences} is $c_{\lfloor n/2 \rfloor} |k|^s {\Lambda^\epsilon \over \epsilon}$, which does contribute a $c_{\lfloor n/2 \rfloor} |k|^s \log \Lambda$ divergence in the $\epsilon \to 0^+$ limit; moreover, in this case, by calculation, $c_{\lfloor n/2 \rfloor} = i_0 + O(\epsilon)$.\footnote{Although we have argued that the hard cutoff prescription $|\ell| \leq \Lambda$ is the natural one to use, it is interesting to note that if instead we impose $|k-\ell| \leq \Lambda$, then still $c_{\lfloor n/2 \rfloor} = i_0 + O(\epsilon)$ when $s = 2\lfloor n/2 \rfloor$ and $\epsilon$ is sufficiently small.}

We can summarize the situation, both for the Archimedean and ultrametric cases, by stating that for $\epsilon = n-s$ positive but sufficiently small, then subject to the restriction that $n$ cannot be a positive even integer when $V = \mathbb{R}^n$,
 \eqn{Iexpand}{
  I_2(k) = \text{(higher powers of $\Lambda$)} + i_2 |k|^s {\Lambda^\epsilon \over \epsilon} 
    + \text{(UV finite)} \,,
 }
where
 \eqn{iTwoCompute}{
  i_2 = \left\{ \seqalign{\span\TR &\qquad\span\TT}{
   i_0 + O(\epsilon) & when $V = \mathbb{R}^n$ and $s$ is a positive even integer  \cr
   0 & otherwise\,.} \right.
 }
The higher powers of $\Lambda$ in \eno{Iexpand} are accompanied by non-negative integer powers of $k^2$, and they correspond to operators which remain relevant in the $\epsilon \to 0^+$ limit.  The only $\log\Lambda$ behavior arising from $I_2(k)$, in the $\epsilon \to 0^+$ limits described above, is the $i_2 |k|^s \log\Lambda$ term coming from the $i_2 |k|^s {\Lambda^\epsilon \over \epsilon}$ term shown in \eno{Iexpand}.  We are not concerned about $O(\epsilon)$ terms to $i_0$ and $i_2$ because they drop out of the $\log\Lambda$ behavior in the $\epsilon \to 0^+$ limit.  In the following sections, therefore, we will drop $O(\epsilon)$ terms from \eno{IzeroAgain} and \eno{iTwoCompute}, and we will evaluate $i_0$ and $i_2$ in terms of $n$ rather than $n_0$.

\section{The propagator through one loop}
\label{TWO}

To derive the tree-level propagator, we use an obvious generalization of \eno{RegConclude} to multi-component scalar fields to rewrite the free action in momentum space:
 \eqn{SfreeAgain}{
  S_2 = {\mu^\epsilon \over 2\hat\gamma}
     g_{ab} \int d^n k \, \hat\phi^a(-k) |k|^s \hat\phi^b(k) \,,
 }
where we recall that $\epsilon = n-s$, and for notational convenience we have introduced\footnote{A point worthy of remark is that while $\hat\gamma$ and $\gamma$ have the same sign in the ultrametric case, and also in the Archimedean case for $0<s<2$, for $2<s<4$ they have the opposite sign.  The integral in \eno{SfreeAgain} is well-defined and positive, so to make our theory sensible we should always choose $\hat\gamma > 0$.  This means that $\gamma < 0$ for $2<s<4$.  As explained in \eno{TwoFourSummary} for a single real scalar, the regulated position space integral used to define the action \eno{Sdef} includes a $(\partial\phi)^2$ term that enters with the opposite sign of the non-local $\left[ \phi(x) - \phi(y) \right]^2$ term, so positivity conditions are difficult to judge in position space.}
 \eqn{GammaHat}{
  \hat\gamma = -{\Gamma_V(n+s) \over 2} \gamma \,.
 }
We immediately extract from \eno{SfreeAgain} the propagator
 \eqn{Ghat}{
  \hat{G}^{ab}(k) &= {\hat\gamma g^{ab} \over \mu^\epsilon |k|^s}  \cr
  G^{ab}(x) &= \Gamma_V(\epsilon) {\hat\gamma g^{ab} \over (\mu |x|)^\epsilon} + 
    \text{(contact terms)} \,.
 }
We are primarily interested in $\epsilon$ small, so that $G^{ab}(x)$ is nearly logarithmic.

Informally, we can understand the one-loop correction to the propagator as a contribution to the 1PI effective action coming from all possible Wick contractions of the two of the four factors of $\phi$ in $S_\text{int}$.  The calculation is done most straightforwardly in momentum space, where we can express
 \eqn{SquarticFourier}{
  S_4 &= -{\mu^\epsilon \over 6\gamma} R_{abcd} \int d^n x \, (\phi^a \phi^c)(x)
    \int' {d^n y \over |x-y|^{n+s}} \left[ (\phi^b \phi^d)(y) - (\phi^b \phi^d)(x) 
      \right]  \cr
   &= -{\mu^\epsilon \over 6\gamma\Gamma_V(n+s)} R_{abcd} \int d^n x \, (\phi^a \phi^c)(x)
     (D^s \phi^b \phi^d)(x)  \cr
   &= {\mu^\epsilon \over 12\hat\gamma} R_{abcd} \int d^n k \, 
    (\widehat{\phi^a \phi^c})(-k) |k|^s (\widehat{\phi^b \phi^d})(k)  \cr
   &= {\mu^\epsilon \over 12\hat\gamma} R_{abcd} 
    \int d^{4n} k \, \delta^n\left( \textstyle{\sum_{i=1}^4} k_i \right) 
     \hat\phi^a(k_1) \hat\phi^b(k_2) \hat\phi^c(k_3)
      \hat\phi^d(k_4) |k_2+k_4|^s \,,
 }
where $d^{4n} k = \prod_{i=1}^4 d^n k_i$.  As usual, derivative terms are implied in $\int'$ in the Archimedean case when $s>2$.  Symbolically, the Wick-contracted quartic action is
 \eqn{SquarticWick}{
  S_4^\text{Wick} &= {\mu^\epsilon \over 12\hat\gamma} R_{abcd} 
    \int d^{4n} k \, \delta^n\left( \textstyle{\sum_{i=1}^4} k_i \right) 
     \hat\phi^a(k_1) 
      \contraction{}{\hat\phi^b}{(k_2)}{\hat\phi^c} \hat\phi^b(k_2) \hat\phi^c(k_3)
      \hat\phi^d(k_4) |k_2+k_4|^s  \cr
   &\hskip2in{} + \text{($\hat\phi^a \hat\phi^d$ contraction)} \,.
 }
We understand $\contraction{}{\hat\phi^b}{(k_2)}{\hat\phi^c} \hat\phi^b(k_2) \hat\phi^c(k_3)$ to mean a replacement of $\hat\phi^b(k_2) \hat\phi^c(k_3)$ by $\hat{G}^{bc}(k_2) \delta^n(k_2+k_3)$.  We omit the $\hat\phi^a\hat\phi^b$ and $\hat\phi^c\hat\phi^d$ contractions from \eno{SquarticWick} because of the antisymmetry of $R_{abcd}$ in $ab$ and $cd$.  We omit the $\hat\phi^a\hat\phi^c$ and $\hat\phi^b\hat\phi^d$ contractions because they include a factor $|k_1+k_3|^s \delta^n(k_1+k_3)$, which vanishes when $s>0$.  After some straightforward algebra, we obtain from \eno{SquarticWick} the form
 \eqn{SWickAgain}{
  S_4^\text{Wick} = 
    -{1 \over 6} R_{ab} \int d^n k \, \hat\phi^a(-k) I_2(k) \hat\phi^b(k) \,,
 }
where $I_2(k)$ is given in \eno{Itwo}.

As discussed below \eno{Itwo}, for suitably small positive $\epsilon=n-s$, $I_2(k)$ includes a term $i_2 |k|^s {\Lambda^\epsilon \over \epsilon}$ iff $V = \mathbb{R}^n$ and $s$ is a positive even integer.  This is the case which leads to local non-linear sigma models.  Otherwise---excluding the case $V = \mathbb{R}^n$ with positive even $n$---the divergent terms in $I_2(k)$ are proportional to $|k|^{2r} \Lambda^{n-2r}$ for non-negative powers $n-2r$ which remain finite as $\epsilon \to 0^+$.  Therefore, apart from the case of local non-linear sigma models, the effects of the ultraviolet divergences in \eno{SWickAgain} are limited to generating relevant, local interactions.  We assume that relevant terms of this type can be tuned away.

It would be tempting at this point to conclude that the non-local action \eno{Sdef} is non-renormalized, as in the case \eno{Skin}.  The reality is more subtle: We will see in section~\ref{FOUR} that higher point diagrams generate one loop divergences that require non-local counterterms; however they are not quite of the form \eno{Sdef}, involving instead the target space laplacian of $Q(X,Y)$.

\section{Curvature and arc length calculations}
\label{DIFFERENTIAL}

Higher point amplitudes in the bi-local non-linear sigma model involve tensors of large rank.  We will therefore find it convenient to introduce some abbreviated notation, based on the following equivalences:
 \eqn{Abbreviations}{
   \begin{tabular}{c||c|c|c}
    standard & $a_1 a_2$ & $\nabla_{(a_1} \nabla_{a_2)}$ & $X^{a_1} X^{a_2}$ \\ \hline
    abbreviated & $a_{12}$ & $\nabla_{a_{12}}$ & $X^{a_{12}}$
   \end{tabular}
 }
Here $(ab) = {1 \over 2} (ab+ba)$.  We employ obvious extensions of \eno{Abbreviations} to larger index sets, e.g.~$X^{a_{123}}$ means $X^{a_1} X^{a_2} X^{a_3}$.

We will often need to simplify expressions involving the curvature tensor and its covariant derivatives.  A primary tool is the Bianchi identities, which we may write using our abbreviated notation as
 \eqn{Bianchi}{
  R_{a_{1234}} + R_{a_{1423}} + R_{a_{1342}} &= 0  \cr
  \nabla_{a_5} R_{a_{1234}} + \nabla_{a_1} R_{a_{2534}} + \nabla_{a_2} R_{a_{5134}} &= 0 \,.
 }
A contracted form of the second Bianchi identity,
 \eqn{dRC}{
  \nabla_b R^b{}_{a_{123}} = \nabla_{a_2} R_{a_{13}} - \nabla_{a_3} R_{a_{12}}
 }
shows that any three-index contraction of $\nabla_{a_5} R_{a_{1234}}$ (meaning any contraction leaving three indices free) can be expressed as linear combinations of re-indexed versions of the tensor $\nabla_{a_1} R_{a_{23}}$; in this sense $\nabla_{a_1} R_{a_{23}}$ by itself is a basis for all the three-index contractions of $\nabla_{a_1} R_{a_{2345}}$.  This observation will be useful to us when we consider the possible Wick contractions of the five-point interaction vertex in the bi-local non-linear sigma model.

Acting on the contracted second Bianchi identity \eno{dRC} with $\nabla_{b_1}$ gives
 \eqn{ddRC}{
  \nabla_{b_1b_2} R^{b_2}{}_{a_{123}} &= \nabla_{b_1a_2} R_{a_{13}} -
    \nabla_{b_1a_3} R_{a_{12}}  \cr
   &\qquad{} + {1 \over 2} [\nabla_{b_1},\nabla_{a_2}] R_{a_{13}} - 
    {1 \over 2} [\nabla_{b_1},\nabla_{a_3}] R_{a_{12}} -
    {1 \over 2} [\nabla_{b_1},\nabla_{b_2}] R^{b_2}{}_{a_{123}} \,.
 }
We describe the terms in the second line of \eno{ddRC} as commutator terms.  Evidently, they can be written as curvature bilinears, meaning contractions of two factors of the Riemann and/or Ricci tensors, with no covariant derivatives.  Acting on the uncontracted second Bianchi identity (the second line of \eno{Bianchi}) with $\nabla^{a_5}$ gives
 \eqn{ddR}{
  \nabla^2 R_{a_{1234}} &= \nabla_{a_{13}} R_{a_{24}} + \nabla_{a_{24}} R_{a_{13}} - 
    \nabla_{a_{14}} R_{a_{23}} - \nabla_{a_{23}} R_{a_{14}} + \text{(commutators)} \,,
 }
where $\nabla^2 = \nabla^b \nabla_b$, and the commutator terms are similar to the ones occurring in \eno{ddRC}: In particular, they are curvature bilinears.  The results \eno{ddRC} and \eno{ddR} show that all four-index contractions of $\nabla_{a_{56}} R_{a_{1234}}$ can be expressed in terms of linear combinations of re-indexed versions of the tensor $\nabla_{a_{12}} R_{a_{34}}$, together with curvature bilinears.

So far, all formulas in this section have been entirely independent of the choice of coordinate system.  We now pass to Riemann normal coordinates in order to study the square of the arc length, $Q(X,Y) = d(X,Y)^2$ between two points $X$ and $Y$. We have from \cite{Brewin:2009se}\footnote{Note however that the results leading to $Q_6^{RR}$ in \cite{Brewin:2009se} contain errors.  In particular, $44$ should have been $4$ in the first line of (11.24).}
 \eqn{ArcLength}{
  Q(X,Y) &= g_{a_{12}} (X^{a_1}-Y^{a_1}) (X^{a_2}-Y^{a_2}) + \sum_{r>3} Q_r(X,Y)  \cr
  Q_4(X,Y) &= -{1 \over 3} R_{a_{1234}} X^{a_{13}} Y^{a_{24}}  \cr
  Q_5(X,Y) &= -{1 \over 12} \nabla_{a_5} R_{a_{1234}} X^{a_{13}} Y^{a_{24}} 
    (X^{a_5} + Y^{a_5})  \cr
  Q_6(X,Y) &= Q_6^{\nabla\nabla R}(X,Y) + Q_6^{RR}(X,Y)  \cr
  Q_6^{\nabla\nabla R}(X,Y) &= -{1 \over 60} \nabla_{a_{56}} R_{a_{1234}}
    (X^{a_{1356}} Y^{a_{24}} + X^{a_{13}} Y^{a_{2456}} + X^{a_{135}} Y^{a_{246}})  \cr
  Q_6^{RR}(X,Y) &= {1 \over 45} R^b{}_{a_{123}} R_{ba_{456}} (4 X^{a_{125}} Y^{a_{346}} - 
   X^{a_{1245}} Y^{a_{36}} - X^{a_{25}} Y^{a_{1346}})
 }
Here $g_{a_{12}}$, $R_{a_{1234}}$, and its derivatives are all evaluated at the origin of Riemann normal coordinates, which is the origin in terms of the coordinates $X^a$ and $Y^a$ used in \eno{ArcLength}.

\def\kTwoInt#1{\int d^{2n} k_{#1} \, \delta^n(k_{#1}) \hat\phi^{a_{#1}}(k_{#1})}
\def\kThreeInt#1{\int d^{3n} k_{#1} \, \delta^n(k_{#1}) \hat\phi^{a_{#1}}(k_{#1})}
\def\kThreeTwoInt#1#2{\int d^{3n} k_{#1} \, \delta^n(k_{#1}) \hat\phi^{a_{#1}}(k_{#1}) \int d^{2n} k_{#2} \, \delta^n(k_{#2})}

\section{Three-point vertices}
\label{THREE}

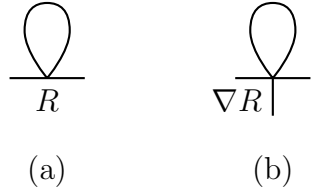
\begin{figure}
  \centerline{\begin{tikzpicture}
   \draw[thick] (-0.5,0) -- (0.5,0);
   \draw[thick] plot [smooth, tension=1] coordinates {(0,0) (-0.3,0.63) (0,1) (0.3,0.63) (0,0)};
   \node[below] at (0,0) {$R$};
   \node[below=25pt] at (0,0) {(a)};
   \draw[thick] (2.5,0) -- (3.5,0);
   \draw[thick] plot [smooth, tension=1] coordinates {(3,0) (2.7,0.63) (3,1) (3.3,0.63) (3,0)};
   \draw[thick] (3,-0.5) -- (3,0);
   \node[below left] at (3,0) {$\nabla R$};
   \node[below=25pt] at (3,0) {(b)};
  \end{tikzpicture}}
  \caption{(a) The one-loop contribution to the propagator.  (b) The one-loop contribution to the 1PI three-point vertex.}\label{LowOrderDiagrams}
 \end{figure}

There are no three-point vertices at tree-level provided we employ Riemann normal coordinates.  As we will explain in this section, three-point vertices appear to be generated at the one-loop level, by the diagram in figure~\ref{LowOrderDiagrams}b; however, they can be absorbed through field redefinition.

As for our discussion in section~\ref{TWO} of the one-loop corrections to the propagator, the one-loop contribution to the three-point function can be obtained efficiently through Wick contractions in the momentum space of the quintic term in the action:
 \eqn{Sfive}{
  S_5 &= {1 \over 2\gamma} \int' {d^n x d^n y \over |x-y|^{n+s}} Q_5(\phi(x),\phi(y))  \cr
    &= {\mu^\epsilon \over 24\hat\gamma} \nabla_{a_5} R_{a_{1234}} 
     \int \left[ \prod_{i=1}^5 d^n k_i \, \hat\phi^{a_i}(k_i) \right]
     \delta^n\left( \textstyle{\sum_{i=1}^5} k_i \right) |k_{24}|^s \,.
 }
High-dimensional Fourier integrals of the type seen in the second line of \eno{Sfive} are common in our calculations, so we have found it useful to introduce some shorthand notation:
 \eqn{MoreAbbreviations}{
   \begin{tabular}{c||c|c|c|c}
    standard & $d^n k_1 d^n k_2$ &
      $\delta^n(k_1+k_2)$ & $|k_1+k_2|$ & $\hat\phi^{a_1}(k_1) \hat\phi^{a_2}(k_2)$ \\ \hline
    abbreviated & $d^{2n} k_{12}$ & 
      $\delta^n(k_{12})$ & $|k_{12}|$ & $\hat\phi^{a_{12}}(k_{12})$
   \end{tabular}
 }
with obvious extensions to larger index sets.  If $A$ is any ordered set of indices, like $123$, and $|A|$ is the number of indices in the set, then the integrals we see most often are of the form
 \eqn{Jdef}{
  {\cal J}_{\phi,k_A}^{a_A}\left[ q(k_A) \right] &\equiv 
   \int \left[ \prod_{i\in A} d^n k_i \, \hat\phi^{a_i}(k_i) \right]
     \delta^n\left( \textstyle{\sum_{i \in A}} k_i \right) q(k_A) 
   \cr &
    = \int d^{|A|n} k_A \, \delta^n(k_A) \hat\phi^{a_A}(k_A) q(k_A) \,,
 }
where $q(k_A)$ is any function of the $k_i$, and the third expression is just a rewriting of the second using the shorthand notation introduced in \eno{MoreAbbreviations}.  Evidently, the ${\cal J}_\phi$ integrals are convergent for reasonable integrands $q$, like powers of norms of sums of momenta, provided the $\phi^a$ are Schwartz-Bruhat functions.  When there is no risk of confusion, we will omit the subscripted $k_A$ and just write ${\cal J}_\phi^{a_A}$.  We can now rewrite \eno{Sfive} as
 \eqn{SfiveAgain}{
  S_5 = {\mu^\epsilon \over 24\hat\gamma} \nabla_{a_5} R_{a_{1234}}
   {\cal J}_\phi^{a_{12345}}[|k_{24}|^s] \,.
 }
As in section~\ref{TWO}, a Wick contraction amounts to a replacement
 \eqn{WickContract}{
  \contraction{}{\hat\phi^{a_i}}{(k_i)}{\hat\phi^{a_j}}
   \hat\phi^{a_i}(k_i) \hat\phi^{a_j}(k_j) \to 
   \hat{G}^{a_{ij}}(k_j) \delta^n(k_{ij}) = {\hat\gamma \over \mu^\epsilon} 
    g^{a_{ij}} \delta^n(k_{ij}) |k_j|^{-s} \,.
 }
It is helpful to note the following examples of Wick contraction:
 \eqn{WickRules}{
  {\cal J}_{\phi,k_{12A}}^{\contraction{}{a_1}{}{a_2}a_1a_2a_A}[|k_{12}|^s] 
   &= \int d^{(|A|+2)n} k_{12A} \, \delta^n(k_{12A}) 
      \contraction{}{\hat\phi^{a_1}}{(k_1)}{\hat\phi^{a_2}}
     \hat\phi^{a_1}(k_1) \hat\phi^{a_2}(k_2) \hat\phi^{a_A}(k_A) |k_{12}|^s  \cr
   &= {\hat\gamma \over \mu^\epsilon} g^{a_{12}}
    \int d^{|A|n} k_A \, \delta^n(k_A) \hat\phi^{a_A}(k_A)
    \int d^{2n} k_{12} \, \delta^n(k_{12}) |k_2|^{-s} |k_{12}|^s = 0  \cr
  {\cal J}_{\phi,k_{12A}}^{\contraction{}{a_1}{}{a_2}a_1a_2a_A}[|k_C|^s] &= 
   {\hat\gamma \over \mu^\epsilon} g^{a_{12}} 
    \int d^{|A|n} k_A \, \delta^n(k_A) \hat\phi^{a_A}(k_A) |k_C|^s
    \int d^{2n} k_{12} \, \delta^n(k_{12}) |k_2|^{-s}  \cr
   &= {\hat\gamma \over \mu^\epsilon} g^{a_{12}} {\cal J}_{\phi,k_A}^{a_A}[|k_C|^s] I_0
    = {\cal J}_{\phi,k_A}^{\contraction{}{a_1}{}{a_2}a_1a_2a_A}[|k_{12C}|^s]  \cr
  {\cal J}_{\phi,k_{12A}}^{\contraction{}{a_1}{}{a_2}a_1a_2a_A}[|k_{1C}|^s]
   &= {\hat\gamma \over \mu^\epsilon} g^{a_{12}} 
    \int d^{|A|n} k_A \, \delta^n(k_A) \hat\phi^{a_A}(k_A) 
    \int d^{2n} k_{12} \, \delta^n(k_{12}) |k_2|^{-s} |k_{1C}|^s  \cr
   &= {\hat\gamma \over \mu^\epsilon} g^{a_{12}} {\cal J}_{\phi,k_A}^{a_A}[I_2(k_C)] \,.
 }
Here $A$ and $C$ are collections of indices, neither including $1$ or $2$, with $C \subset A$, and we assume $s>0$ in order to obtain the vanishing of the first integral.  Recall from section~\ref{DIVERGENCES} that $I_0$ diverges as $\Lambda^\epsilon/\epsilon$ when $\epsilon = n-s$ is sufficiently small and positive, giving rise to $\log\Lambda$ behavior in the limit $\epsilon \to 0^+$. 

There are ${5 \choose 2} = 10$ possible single Wick contractions of $S_5$, but the $(12)$ contraction (meaning the contraction of $a_1$ and $a_2$) vanishes because $g^{a_{12}} \nabla_{a_5} R_{a_{1234}} = 0$; likewise the $(34)$ contraction vanishes.  Meanwhile the $(14)$ and $(23)$ contractions are equal because both $\nabla_{a_5} R_{a_{1234}}$ and $|k_{24}|^s$ are symmetrical under the simultaneous exchange of $1 \leftrightarrow 3$ and $2 \leftrightarrow 4$.  For the same reason, the $(15)$ and $(35)$ contractions are equal, and so are the $(25)$ and $(45)$ contractions.  Finally, the $(24)$ contraction vanishes because of the first line of \eno{WickRules}.  We are left with
 \eqn{SWickPlan}{
  S_5^\text{Wick} = S_{5,(13)}^\text{Wick} + 2 S_{5,(14)}^\text{Wick} + 
    2 S_{5,(15)}^\text{Wick} + 2 S_{5,(25)}^\text{Wick}
 }
where
 \eqn{SWTerms}{
  S_{5,(13)}^\text{Wick} &= {\mu^\epsilon / \hat\gamma \over 24} \nabla_{a_5} R_{a_{1234}}
    {\cal J}_\phi^{\contraction{}{a_1}{a_2}{a_3} a_1 a_2 a_3 a_{45}}[|k_{24}|^s]
    = {1 \over 24} \nabla_{a_5} R_{a_{24}} {\cal J}_\phi^{a_{245}}[|k_{24}|^s]  \cr
   &= {1 \over 24} \nabla_{a_3} R_{a_{12}} {\cal J}_\phi^{a_{123}}[|k_3|^s] I_0  \cr
  S_{5,(14)}^\text{Wick} &= {\mu^\epsilon / \hat\gamma \over 24} \nabla_{a_5} R_{a_{1234}}
    {\cal J}_\phi^{\contraction{}{a_1}{a_{23}}{a_4} a_1 a_{23} a_4 a_5}[|k_{24}|^s]
    = -{1 \over 24} \nabla_{a_5} R_{a_{23}} {\cal J}_\phi^{a_{235}}[I_2(k_2)]  \cr
   &= -{1 \over 24} \nabla_{a_1} R_{a_{23}} {\cal J}_\phi^{a_{123}}[I_2(k_3)]  \cr
  S_{5,(15)}^\text{Wick} &= {\mu^\epsilon / \hat\gamma \over 24} \nabla_{a_5} R_{a_{1234}}
    {\cal J}_\phi^{\contraction{}{a_1}{a_{234}}{a_5} a_1 a_{234} a_5}[|k_{24}|^s]
   = {1 \over 24} (\nabla_{a_3} R_{a_{24}} - \nabla_{a_4} R_{a_{23}})
     {\cal J}_\phi^{a_{234}}[|k_{24}|^s] I_0 \cr
   &= {1 \over 24} (\nabla_{a_3} R_{a_{12}} - \nabla_{a_1} R_{a_{23}})
     {\cal J}_\phi^{a_{123}}[|k_3|^s] I_0  \cr
  S_{5,(25)}^\text{Wick} &= {\mu^\epsilon / \hat\gamma \over 24} \nabla_{a_5} R_{a_{1234}}
    {\cal J}_\phi^{a_1 \contraction{}{a_2}{a_{34}}{a_5} a_2 a_{34} a_5}[|k_{24}|^s]
   = -{1 \over 24} (\nabla_{a_3} R_{a_{14}} - \nabla_{a_4} R_{a_{13}})
     {\cal J}_\phi^{a_{134}}[I_2(k_4)]  \cr
   &= {1 \over 24} (\nabla_{a_3} R_{a_{12}} - \nabla_{a_1} R_{a_{23}})
     {\cal J}_\phi^{a_{123}}[I_2(k_3)] \,.
 }

A few comments are in order:
 \begin{itemize}
  \item Because $S_{5,(14)}^\text{Wick}$ and $S_{5,(25)}^\text{Wick}$ are proportional to the $I_2$ loop integral, they do not contribute logarithmic divergences in the $\epsilon \to 0^+$ limits described in section~\ref{DIVERGENCES}, except when $s$ is a positive even integer.  In a position space account, these non-logarithmic terms correspond to contractions of $\phi(x)$ with $\phi(y)$.
  \item We are mostly interested in $S_{5,(13)}^\text{Wick}$ and $S_{5,(15)}^\text{Wick}$ because $I_0$ does produce a logarithmic divergence in the $\epsilon \to 0^+$ limit.  Note that in these terms, the vertex factor $|k_{13}|^s$ involves only external momenta.  This is the crucial feature, noted already in the introduction, which allows non-local counterterms to arise.  In position space, the logarithmic terms correspond to contractions of $\phi(y)$ with itself.
 \end{itemize}
In summary,
 \eqn{SWsummary}{
  S_5^\text{Wick} &= {1 \over 24} \kThreeInt{123} \Big[ I_2(k_3) (2 \nabla_{a_3} R_{a_{12}} 
   - 4 \nabla_{a_1} R_{a_{23}})
  \cr &\qquad\qquad\qquad{} 
  + I_0 |k_3|^s (3 \nabla_{a_3} R_{a_{12}} - 2 \nabla_{a_1} R_{a_{23}}) \Big] \,.
 }
The minimal counterterm needed to cancel the $\Lambda^\epsilon/\epsilon$ divergences in \eno{SWsummary} is
 \eqn{SThreeCT}{
  S_3^\text{ct} = {\Lambda^\epsilon \over 2\epsilon} \kThreeInt{123} |k_3|^s 
    \left[ K_1 \nabla_{a_3} R_{a_{12}} + K_2 \nabla_{(a_1} R_{a_2)a_3} \right]
 }
where
 \eqn{KOneTwo}{
  K_1 = -{3i_0+2i_2 \over 12} \qquad\qquad
  K_2 = {i_0+2i_2 \over 6} \,,
 }
and $i_0$ and $i_2$ are as defined in \eqref{IzeroAgain} and \eqref{iTwoCompute}.  We use the notation $S_3^\text{ct}$ to denote a counterterm that is cubic in the fields.  Our primary interest is in cases where $i_2=0$---namely, cases in which $V = \mathbb{Q}_{p^n}$, or $V = \mathbb{R}^n$ but $s$ is not a positive even integer.  However, tracking $I_0$, $I_2(k)$, $i_0$, and $i_2$ throughout our computations is useful as a bookkeeping device in order to simultaneously treat the local and bi-local theories, in the $\epsilon \to 0^+$ limit, with $i_2$ entering in only to describe the local theories.

\section{Renormalization through cubic order in the fields}
\label{RENORMALIZATION}

Before entering into the more complicated story of quartic terms in the action, let's preview the endgame of our analysis, in which we produce a bare action which incorporates the renormalized action plus the counterterms in a form that we can express entirely in terms of arc length.

First, let's rephrase the Wick-contracted quartic action \eno{SWickAgain} as
 \eqn{SWA}{
  S_4^\text{Wick} = -{1 \over 6} \int d^{2n} k_{12} \, \delta^n(k_{12}) \hat\phi^{a_{12}}(k_{12})
    I_2(k_2) R_{a_{12}} \,,
 }
from which we conclude that we need a counterterm quadratic in the fields of the form
 \eqn{STwoCT}{
  S_2^\text{ct} = {\Lambda^\epsilon \over 2\epsilon} 
    \int d^{2n} k_{12} \, \delta^n(k_{12})
    \hat\phi^{a_{12}}(k_{12}) |k_2|^s K_0 R_{a_{12}} \,,
 }
where
 \eqn{KZero}{
  K_0 = {i_2 \over 3} \,.
 }
The results \eno{STwoCT} and \eno{SThreeCT} together put some constraints on the renormalization procedure, but as we will see they do not completely determine it.

The question of renormalizability is whether we can reorganize the renormalized action plus counterterms into a bare action whose form is the same that we started with:
 \eqn{Scompare}{
  S[\phi] + S^\text{ct}[\phi] = S^\text{B}[\phi_\text{B}] \,.
 }
where
 \eqn{SctSum}{
  S^\text{ct}[\phi] = \sum_{r>1} S_r^\text{ct}
 }
is the sum of all counterterms, $S[\phi]$ is as given in \eno{Sdef}, and
 \eqn{SBdef}{
  S^\text{B}[\phi_\text{B}] = 
    {\Lambda^{n-s} \over 2\gamma} \int'_{V \times V} {d^n x d^n y \over |x-y|^{n+s}} 
    Q^\text{B}(\phi_\text{B}(x),\phi_\text{B}(y)) \,,
 }
where $Q^\text{B}(X_\text{B},Y_\text{B}) = d^\text{B}(X_\text{B},Y_\text{B})^2$.  The bare arc length $d^\text{B}$, derived from a bare metric tensor $g^\text{B}_{ab}$, may differ from the renormalized arc length $d$, and the bare coordinates $\phi^a_\text{B}$ may likewise differ from the renormalized coordinates $\phi^a$.  We require however that the points $\phi_\text{B}=0$ and $\phi=0$ coincide.  At tree level, where we ignore all counterterms, we have the relations
 \eqn{TreeLevelRelations}{
  \left( {\Lambda \over \mu} \right)^\epsilon g^\text{B}_{ab}(\phi_\text{B}) = 
    g_{ab}(\phi) \qquad\qquad 
    \phi_\text{B}^a = \phi^a \,,
 }
and our key task is to find perturbative corrections to these relations that render \eno{Scompare} correct.

To begin, let's examine the quadratic terms in \eno{Scompare}, using the counterterm $S_2^\text{ct}$ from \eno{STwoCT}:
 \eqn{QuadraticCompare}{
  &\int d^{2n} k_{12} \, \delta^n(k_{12}) \hat\phi^{a_{12}}(k_{12}) |k_2|^s
    \left[ g_{a_{12}} + {\hat\gamma \over \epsilon} 
     \left( {\Lambda \over \mu} \right)^\epsilon 
      K_0 R_{a_{12}} \right] + O(\phi^3) + O(\hat\gamma^2)
    \cr &\qquad\qquad{}
   = \left( {\Lambda \over \mu} \right)^\epsilon \int d^{2n} k_{12} \, \delta^n(k_{12}) 
     \hat\phi_\text{B}^{a_{12}}(k_{12}) |k_2|^s g^\text{B}_{a_{12}} \,,
 }
where for simplicity we multiplied through by $2\hat\gamma/\mu^\epsilon$. We surmise from \eno{QuadraticCompare} that corrections to \eno{TreeLevelRelations} can be expressed as a power series in the dimensionless quantity
 \eqn{GammaTilde}{
  \tilde\gamma = {\hat\gamma \over \epsilon} \left( {\Lambda \over \mu} \right)^\epsilon \,.
 }
That is, 
 \eqn{CoordinateMap}{
  \phi_\text{B}^a(x) = \phi^a(x) 
    + \tilde\gamma \left[ U^a{}_b \phi^b(x) 
    + V^a{}_{b_{12}} \phi^{b_{12}}(x)
    + W^a{}_{b_{123}} \phi^{b_{123}}(x) \right]
    + O(\phi^4) + O(\tilde\gamma^2) \,,
 }
where $V^a{}_{b_{12}}$ and $W^a{}_{b_{123}}$ are fully symmetric in their lower indices, and $U$, $V$, and $W$ are all independent of $\phi$ (and $\phi_\text{B}$).  In other words, \eno{CoordinateMap} is a Taylor expansion of $\phi_\text{B}^a$ in the coordinates $\phi^a$.  Also,
 \eqn{gBare}{
  \left( {\Lambda \over \mu} \right)^\epsilon g^\text{B}_{ab}(\phi_\text{B}) = g_{ab}(\phi) + 
    \tilde\gamma T_{ab}(\phi) + O(\tilde\gamma^2)
 }
for some tensor $T_{ab}(\phi)$.  As our notation indicates, $T_{ab}(\phi)$ does depend on $\phi$.  As with other tensors, if we omit the argument, we mean that $T_{ab}$ is evaluated at $\phi=0$.  
Using \eno{CoordinateMap} and \eno{gBare}, we see that \eno{QuadraticCompare} is satisfied provided
 \eqn{MetricCompare}{
  g_{a_{12}} + \tilde\gamma K_0 R_{a_{12}}
   = (\delta^{b_1}_{a_1} + \tilde\gamma U^{b_1}{}_{a_1})
     (\delta^{b_2}_{a_2} + \tilde\gamma U^{b_2}{}_{a_2})
     (g_{b_{12}} + \tilde\gamma T_{b_{12}}) + O(\hat\gamma^2) \,,
 }
or in other words provided
 \eqn{RST}{
  T_{a_{12}} + 2U_{a_{(12)}} = K_0 R_{a_{12}} \,,
 }
where
 \eqn{Slower}{
  U_{a_{12}} = U^b{}_{a_2} g_{a_1b} \qquad\qquad
  U_{a_{(12)}} = {1 \over 2} (U_{a_{12}} + U_{a_{21}}) \,.
 }
It should be kept in mind that $T_{a_{12}}$ is the $\phi=0$ value of a tensor field $T_{a_{12}}(\phi)$ defined over the whole of $M$, whereas $U_{a_{(12)}}$ is defined only at $\phi=0$.  Let's assume that
 \eqn{STassume}{
  T_{a_{12}}(\phi) = t_0 R_{a_{12}}(\phi) \qquad\qquad U_{a_{(12)}} = u_0 R_{a_{12}} \,.
 }
(A term in $T_{a_{12}}(\phi)$ proportional to $R(\phi) g_{a_{12}}(\phi)$ is also possible, but the divergences we will encounter do not require it.)  Then \eno{RST} reduces to
 \eqn{stiConstraint}{
  t_0 + 2u_0 = K_0 \,.
 }
As previously noted, based on the treatment of quadratic terms alone, we cannot distinguished between metric renormalization (related to the coefficient $t_0$) and field redefinition (related to the coefficient $u_0$).

In order to proceed to higher orders, we require the squared arc length formula for bare quantities:
 \eqn{BareArc}{
  Q^\text{B}(X_\text{B},Y_\text{B}) = 
    g^\text{B}_{a_{12}} (X_\text{B}^{a_1} - Y_\text{B}^{a_1}) 
     (X_\text{B}^{a_2} - Y_\text{B}^{a_2}) + 
      \sum_{r>2} Q^\text{B}_r(X_\text{B},Y_\text{B}) \,.
 }
We do not require $\phi_\text{B}$ to be Riemann normal coordinates for $d^\text{B}$, so there are contributions to $Q^\text{B}$ at cubic order:
 \eqn{CubicQ}{
  Q^\text{B}_3(X_\text{B},Y_\text{B}) = \Gamma^\text{B}_{a_{123}} 
   (X_\text{B}^{a_1}-Y_\text{B}^{a_1}) (X_\text{B}^{a_2}-Y_\text{B}^{a_2})
    (X_\text{B}^{a_3}+Y_\text{B}^{a_3}) \,,
 }
where $\Gamma^\text{B}_{abc} = g^\text{B}_{ad} \Gamma^{\text{B}d}_{bc}$ and $\Gamma^{\text{B}a}_{bc}$ is the Christoffel connection for $g^\text{B}_{ab}$.  From \eno{gBare} we have immediately 
 \eqn{GammaB}{
  \left( {\Lambda \over \mu} \right)^\epsilon \Gamma^\text{B}_{a_{312}} = 
    \tilde\gamma \left[ \nabla_{(a_1} T_{a_2)a_3} - 
     {1 \over 2} \nabla_{a_3} T_{a_{12}} \right] + O(\tilde\gamma^2) \,,
 }
From $Q^\text{B}_3$ we obtain a cubic term in the bare action:
 \eqn{CubicS}{
  S^\text{B}_3 &= {\Lambda^\epsilon \over 2\gamma} 
        \int' {d^n x d^n y \over |x-y|^{n+s}} Q^\text{B}_3(\phi_\text{B}(x),\phi_\text{B}(y))
    = {\Lambda^\epsilon \over 2\hat\gamma} 
        {\cal J}_{\phi_\text{B}}^{a_{123}}[|k_3|^s] \Gamma^\text{B}_{a_{312}}  \cr
   &= {\Lambda^\epsilon \over 2\epsilon} {\cal J}_\phi^{a_{123}}[|k_3|^s]
     \left[ \nabla_{(a_1} T_{a_2)a_3} - {1 \over 2} \nabla_{a_3} T_{a_{12}} \right] + 
       O(\tilde\gamma) \,.
 }
Another term cubic in $\phi$ arises in the bare action from plugging the non-linear field redefinition \eno{CoordinateMap} into the quadratic term $S^\text{B}_2$.  To work this out, it helps first to note that passing \eno{CoordinateMap} through a Fourier transform yields
 \eqn{MomentumMap}{
  \hat\phi_\text{B}^a(k) = \hat\phi^a(k) + \tilde\gamma \delta\hat\phi^a(k) + O(\tilde\gamma^2)
 }
where
 \eqn{deltaHatPhi}{
  \delta\hat\phi^a(k) = 
    U^a{}_b \hat\phi^b(k) + V^a{}_{b_{12}} (\hat\phi^{b_1} * \hat\phi^{b_2})(k) + 
    W^a{}_{b_{123}} (\hat\phi^{b_1} * \hat\phi^{b_2} * \hat\phi^{b_3})(k) +
    O(\phi^4)
 }
and $*$ denotes convolution.  It follows immediately that
 \eqn{JtwoExpand}{
  {\cal J}_{\phi_\text{B}}^{a_{12}}[|k_2|^s] &= 
   {\cal J}_\phi^{a_{12}}[|k_2|^s] +
    2\tilde\gamma 
     \int d^{2n} k_{12} \, \delta^n(k_{12}) \delta\hat\phi^{(a_1}(k_1) 
       \hat\phi^{a_2)}(k_2) |k_2|^s + O(\phi^5) + O(\tilde\gamma^2)  \cr
   &= {\cal J}_\phi^{a_{12}}[|k_2|^s] + 
      2\tilde\gamma \Big[ U^{(a_1}{}_b {\cal J}_{\phi,k_2\ell}^{a_2)b}[|k_2|^s] + 
      V^{(a_1}{}_{b_{12}} {\cal J}_{\phi,k_2\ell_{12}}^{a_2)b_{12}}[|k_2|^s]
       \cr &\qquad\qquad{}
      + W^{(a_1}{}_{b_{123}} {\cal J}_{\phi,k_2\ell_{123}}^{a_2)b_{123}}[|k_2|^s] \Big] + 
       O(\phi^5) + O(\tilde\gamma^2) \,,
 }
and so
 \eqn{StwoExpand}{
  S^\text{B}_2 = {\Lambda^\epsilon \over 2\hat\gamma} 
   {\cal J}_{\phi_\text{B}}^{a_{12}}[|k_2|^s] g^\text{B}_{a_{12}}
  &= S_2 + {\Lambda^\epsilon \over 2\epsilon} {\cal J}_\phi^{a_{12}}[|k_2|^s]
    \left[ T_{a_{12}} + 2 U_{a_{12}} \right] + 
    {\Lambda^\epsilon \over \epsilon} {\cal J}_\phi^{a_{123}}[|k_3|^s] V_{a_{312}}
    \cr &\qquad\qquad{}
   + {\Lambda^\epsilon \over \epsilon} {\cal J}_\phi^{a_{1234}}[|k_4|^s] W_{a_{4123}} + 
    O(\phi^5) + O(\tilde\gamma) \,,
 }
where we are lowering indices on $V$ and $W$ with the renormalized metric $g_{ab}$.  The $T_{a_{12}}+2U_{a_{12}}$ term in \eno{StwoExpand} is the same combination we saw in \eno{RST}, with the symmetrization $U_{a_{12}} \to U_{a_{(12)}}$  implied because we multiply by ${\cal J}_\phi^{a_{12}}[|k_2|^s]$, which is symmetric.  The next term in \eno{StwoExpand} is the interesting one for us.  The only constraint on $V_{a_{312}}$ is symmetry in the $12$ indices.  This is the same symmetry that ${\cal J}_\phi^{a_{123}}[|k_3|^s]$ possesses.  Therefore ${\cal J}_\phi^{a_{123}}[|k_3|^s] V_{a_{312}}$ is the most general linear combination of terms coming from ${\cal J}_\phi^{a_{123}}[|k_3|^s]$ integrals.  Likewise, the only constraint on $W_{a_{4123}}$ is symmetry in $123$, so the last term shown explicitly in \eno{StwoExpand} is the most general linear combination of terms coming from ${\cal J}_\phi^{a_{1234}}[|k_4|^s]$ integrals.

We now have all the ingredients needed to calculate the $O(\phi^3)$ correction to \eno{QuadraticCompare}.  Specifically, we expand \eno{Scompare} to cubic order in the renormalized fields, using the expression \eno{SThreeCT} for $S^\text{ct}_3$, as well as $S^\text{B}_3$ from \eno{CubicS} and the $O(\phi^3)$ term from \eno{StwoExpand}.  The result is
 \eqn{CubicCompare}{
  {\cal J}_\phi^{a_{123}}[|k_3|^s] &
    \left[ K_1 \nabla_{a_3} R_{a_{12}} + K_2 \nabla_{(a_1} R_{a_2)a_3} \right]  
    \cr &
    = {\cal J}_\phi^{a_{123}}[|k_3|^s]
    \left[ \nabla_{(a_1} T_{a_2)a_3} - {1 \over 2} \nabla_{a_3} T_{a_{12}} + 
        2V_{a_{312}} \right] \,.
 }
Evidently, we may set
 \eqn{UParams}{
  V_{a_{312}} = v_1 \nabla_{a_3} R_{a_{12}} + v_2 \nabla_{(a_1} R_{a_2)a_3} \,,
 }
where
 \eqn{utiConstraints}{
  -{1 \over 2} t_0 + 2 v_1 = K_1  \qquad\qquad
  t_0 + 2 v_2 = K_2 \,.
 }
The larger message is that $V_{a_{312}}$ is sufficiently unconstrained that we could use it to absorb any counterterm proportional to ${\cal J}_\phi^{a_{123}}[|k_3|^s]$.  By the same token, when we get to quartic order, the field redefinition coefficients $W_{a_{4123}}$ can be used to absorb any terms proportional to ${\cal J}_\phi^{a_{1234}}[|k_4|^s]$.  Therefore, when we do proceed to quartic order, we may simplify our work by systematically dropping all such terms.  We will see in section~\ref{FOUR} that other terms emerge, proportional to ${\cal J}_\phi^{a_{1234}}[|k_{24}|^s]$.  These are the ones that cannot be absorbed into field redefinitions.

\section{Quartic counterterms}
\label{FOUR}

 \begin{figure}
  \centerline{\begin{tikzpicture}
   \draw[thick] (-1/2,0) -- (1/2,0);
   \draw[thick] (-0.25,-0.43) -- (0,0) -- (0.25,-0.43);
   \draw[thick] plot [smooth, tension=1] coordinates {(0,0) (-0.2,0.42) (0,0.67) (0.2,0.42) (0,0)};
   \node[below left] at (-0.2,0) {$\nabla\nabla R$};
   \node[below=25pt] at (0,0) {(a)};
   \draw[thick] (2.5,0) -- (3.5,0);
   \draw[thick] (2.75,-0.43) -- (3,0) -- (3.25,-0.43);
   \draw[thick] plot [smooth, tension=1] coordinates {(3,0) (2.8,0.42) (3,0.67) (3.2,0.42) (3,0)};
   \node[below left] at (2.8,0) {$RR$};
   \node[below=25pt] at (3,0) {(b)};
   \draw[thick] (5.65,0.35) -- (6,0) -- (5.65,-0.35);
   \draw[thick] (7.35,0.35) -- (7,0) -- (7.35,-0.35);
   \draw[thick] plot [smooth, tension=1] coordinates {(6,0) (6.5,0.4) (7,0)};
   \draw[thick] plot [smooth, tension=1] coordinates {(6,0) (6.5,-0.4) (7,0)};
   \node[left=5pt] at (6,0) {$R$};
   \node[right=5pt] at (7,0) {$R$};
   \node[below=25pt] at (6.5,0) {(c)};
  \end{tikzpicture}}
  \caption{One-loop contributions to the 1PI four-point vertex: (a) Single Wick contractions of the $\nabla\nabla R$ six-point vertices; (b) Single Wick contractions of $RR$ six-point vertices; (c) Diagrams involving only four-point vertices.}\label{QuarticDiagrams}
 \end{figure}
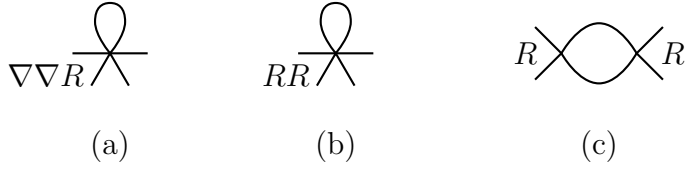
Four-point vertices are present at tree-level, and they are also generated by three different types of one-loop diagrams, as shown in figure~\ref{QuarticDiagrams}.  Our goal in this section is to evaluate one-loop divergences proportional to $\nabla_{a_{12}} R_{a_{34}}$ (and re-indexings of it).  The claim is that only the diagram in figure~\ref{QuarticDiagrams}a can contribute.  Tracking only target space indices, the vertex factor in this diagram is $\nabla_{a_{56}} R_{a_{1234}}$, and the internal propagator can tie any two of the six indices together.  So the diagram as a whole must be proportional to four-index contractions of $\nabla_{a_{56}} R_{a_{1234}}$.  As explained in section~\ref{DIFFERENTIAL}, re-indexed versions of $\nabla_{a_{12}} R_{a_{34}}$, together with curvature bilinears, provide a basis for such contractions.  The diagrams in figure~\ref{QuarticDiagrams}b and~\ref{QuarticDiagrams}c are manifestly proportional to curvature bilinears, so they cannot contribute terms proportional to $\nabla_{a_{12}} R_{a_{34}}$.  (One immediate way to see this is that the Riemann tensor could vanish at $\phi=0$ without its derivatives vanishing.)  In the explicit calculations below, we will encounter and discard many curvature bilinear terms, which we generically write as $O(RR)$, meaning some contraction of $R_{a_{1234}} R_{a_{5678}}$.

We claim that the counterterms needed to cancel the divergences from the diagram in figure~\ref{QuarticDiagrams}a proportional to $\nabla_{a_{12}} R_{a_{34}}$ (and re-indexings of it) are 
 \eqn{GradGradSummary}{
  S_{6,\nabla\nabla R}^\text{ct} &= {\Lambda^\epsilon \over 2\epsilon} \Big[
    {\cal J}_\phi^{a_{1234}}[|k_{24}|^s] \left( 
     K_3 {\color{red} \nabla_{a_{12}} R_{a_{34}}} + 
     K_4 {\color[rgb]{0,0.6,0} \nabla_{a_{13}} R_{a_{24}}} \right)
  \cr&\qquad\qquad\qquad{}
    + {\cal J}_\phi^{a_{1234}}[|k_4|^s] \left(
     K_5 {\color{brown} \nabla_{a_{12}} R_{a_{34}}} + 
     K_6 {\color{blue} \nabla_{a_{34}} R_{a_{12}}} \right)
   \Big] \,,
 }
where
 \eqn{KValues}{\seqalign{\span\TC}{
  K_3 = {3 i_0 + 2 i_2 \over 30}  \qquad\qquad
  K_4 = -{7 i_0 + 3 i_2 \over 60}  \cr
  K_5 = {i_0 + 3 i_2 \over 30}  \qquad\qquad
  K_6 = -{3 i_0 + 4i_2 \over 60} \,.
 }}
Color-coding in \eno{GradGradSummary} is to help track to which term in \eno{GradGradSummary} each of the many terms in later equations contribute.  The remainder of this section is devoted to deriving \eno{KValues}.

To derive the sixth-order vertex used in figure~\ref{QuarticDiagrams}a, we start from the $Q_6^{\nabla\nabla R}$ term in \eno{ArcLength} and extract the following six-order terms in the action:
 \eqn{SGradGrad}{
  S_{6,\nabla\nabla R} &= {\mu^\epsilon \over 2\gamma} \int' {d^n x d^n y \over |x-y|^{n+s}}
     Q_6^{\nabla\nabla R}(\phi(x),\phi(y))
    = S_{6,\nabla\nabla R}^{4+2} + S_{6,\nabla\nabla R}^{3+3}
 }
where
 \eqn{SSixSplit}{
  S_{6,\nabla\nabla R}^{4+2} &= {\mu^\epsilon \over 120\hat\gamma} \nabla_{a_{56}} R_{a_{1234}} 
     {\cal J}_\phi^{a_{123456}}[|k_{24}|^s]  \cr
  S_{6,\nabla\nabla R}^{3+3} &= {\mu^\epsilon \over 240\hat\gamma} \nabla_{a_{56}} R_{a_{1234}} 
     {\cal J}_\phi^{a_{123456}}[|k_{246}|^s] \,.
 }
Our task is to compute the counterterms for all the single Wick contractions of $S_{6,\nabla\nabla R}^{4+2}$ and $S_{6,\nabla\nabla R}^{3+3}$.

For $S_{6,\nabla\nabla R}^{4+2}$, relations among Wick contractions that are obvious from symmetries plus the first line of \eno{WickRules} are as follows:
 \eqn{FourTwoRelations}{\seqalign{\span\TC}{
  (12) = 0 \qquad (34) = 0 \qquad (24) = 0  \cr
  (14) = (23) \qquad (15) = (35) = (16) = (36) \qquad
  (25) = (45) = (26) = (46) \,.
 }}
Thus, of fifteen single Wick contractions of $S_{6,\nabla\nabla R}^{4+2}$, there are actually only five that determine the full answer:
 \eqn{SFourTwoPlan}{
  S_{6,\nabla\nabla R}^{4+2,\text{Wick}} 
   = S_{6,\nabla\nabla R,(13)}^{4+2,\text{Wick}}
     + 2 S_{6,\nabla\nabla R,(14)}^{4+2,\text{Wick}}
     + 4 S_{6,\nabla\nabla R,(15)}^{4+2,\text{Wick}}
     + 4 S_{6,\nabla\nabla R,(25)}^{4+2,\text{Wick}}
     + S_{6,\nabla\nabla R,(56)}^{4+2,\text{Wick}} \,.
 }
If we write the counterterm for an expression $Q$ as $\{Q\}_\text{ct}$, then the rules of computation we need are a trivial adaptation of \eno{WickRules}:
 \eqn{CountertermRules}{
  \left\{ {\cal J}_\phi^{\contraction{}{a_1}{}{a_2}a_1a_2a_A}[|k_C|^s] \right\}_\text{ct} &= 
  \left\{ {\cal J}_\phi^{\contraction{}{a_1}{}{a_2}a_1a_2a_A}[|k_{12C}|^s] \right\}_\text{ct}
   = -\tilde\gamma g^{a_{12}} {\cal J}_\phi^{a_A}[|k_C|^s] i_0  \cr
  \left\{ {\cal J}_\phi^{\contraction{}{a_1}{}{a_2}a_1a_2a_A}[|k_{1C}|^s] \right\}_\text{ct}
   &= -\tilde\gamma g^{a_{12}} {\cal J}_\phi^{a_A}[|k_C|^s] i_2 \,.
 }
The counterterms that we need to cancel divergences coming from the single Wick contractions of $S_{6,\nabla\nabla R}^{4+2}$, as shown in \eno{SFourTwoPlan}, are
 \eqn{SctPlan}{
  S_{6,\nabla\nabla R}^{4+2,\text{ct}} = {\Lambda^\epsilon \over 120\epsilon} \left[
   I_{6,\nabla\nabla R,(13)}^{4+2,\text{ct}}
     + 2 I_{6,\nabla\nabla R,(14)}^{4+2,\text{ct}}
     + 4 I_{6,\nabla\nabla R,(15)}^{4+2,\text{ct}}
     + 4 I_{6,\nabla\nabla R,(25)}^{4+2,\text{ct}}
     + I_{6,\nabla\nabla R,(56)}^{4+2,\text{ct}}
   \right]
 }
where
 \eqn{FourTwoCounterterms}{
  I_{6,\nabla\nabla R,(13)}^{4+2,\text{ct}} &= {1 \over \tilde\gamma}
    \nabla_{a_{56}} R_{a_{1234}}
     \left\{ {\cal J}^{\contraction{}{a_1}{a_2}{a_3} a_1 a_2 a_3 a_{456}}_\phi[|k_{24}|^s]
       \right\}_\text{ct}
    = -\nabla_{a_{56}} R_{a_{24}} {\cal J}_\phi^{a_{2456}}[|k_{24}|^s] i_0  \cr
   &= -{\color[rgb]{0,0.6,0} \nabla_{a_{13}} R_{a_{24}}} {\cal J}_\phi^{a_{1234}}[|k_{24}|^s] i_0  \cr
  I_{6,\nabla\nabla R,(14)}^{4+2,\text{ct}} &= {1 \over \tilde\gamma}
    \nabla_{a_{56}} R_{a_{1234}} 
     \left\{ {\cal J}^{\contraction{}{a_1}{a_{23}}{a_4} a_1 a_{23} a_4 a_{56}}_\phi[|k_{24}|^s]
       \right\}_\text{ct}
    = \nabla_{a_{56}} R_{a_{23}} {\cal J}_\phi^{a_{2356}}[|k_2|^s] i_2  \cr
   &= {\color{brown} \nabla_{a_{12}} R_{a_{34}}} {\cal J}_\phi^{a_{1234}}[|k_4|^s] i_2  \cr
  I_{6,\nabla\nabla R,(15)}^{4+2,\text{ct}} &= {1 \over \tilde\gamma}
    \nabla_{a_{56}} R_{a_{1234}}
     \left\{ {\cal J}^{\contraction{}{a_1}{a_{234}}{a_5} a_1 a_{234} a_5 a_6}_\phi[|k_{24}|^s]
       \right\}_\text{ct}
    = -\nabla_{ba_6} R^b{}_{a_{234}} {\cal J}_\phi^{a_{2346}}[|k_{24}|^s] i_0  \cr
   &= (-\nabla_{a_{63}} R_{a_{24}} + \nabla_{a_{64}} R_{a_{23}}) 
        {\cal J}_\phi^{a_{2346}}[|k_{24}|^s] i_0 + O(RR)  \cr
   &= ({\color{red} \nabla_{a_{12}} R_{a_{34}}} - {\color[rgb]{0,0.6,0} \nabla_{a_{13}} R_{a_{24}}}) 
        {\cal J}_\phi^{a_{1234}}[|k_{24}|^s] i_0 + O(RR)  \cr
  I_{6,\nabla\nabla R,(25)}^{4+2,\text{ct}} &= {1 \over \tilde\gamma}
    \nabla_{a_{56}} R_{a_{1234}}
     \left\{ {\cal J}^{a_1 \contraction{}{a_2}{a_{34}}{a_5} a_2 a_{34} a_5 a_6}_\phi[|k_{24}|^s]
       \right\}_\text{ct}
    = -\nabla_{ba_6} R_{a_1}{}^b{}_{a_{34}} {\cal J}_\phi^{a_{1346}}[|k_4|^s] i_2  \cr
    &= (\nabla_{a_{63}} R_{a_{14}} - \nabla_{a_{64}} R_{a_{13}}) 
        {\cal J}_\phi^{a_{1346}}[|k_4|^s] i_2 + O(RR)  \cr
    &= ({\color{brown} \nabla_{a_{12}} R_{a_{34}}} - {\color{blue} \nabla_{a_{34}} R_{a_{12}}}) 
        {\cal J}_\phi^{a_{1234}}[|k_4|^s] i_2 + O(RR)  \cr
  I_{6,\nabla\nabla R,(56)}^{4+2,\text{ct}} &= {1 \over \tilde\gamma}
    \nabla_{a_{56}} R_{a_{1234}}
     \left\{ {\cal J}^{a_{1234} \contraction{}{a_5}{}{a_6} a_5 a_6}_\phi[|k_{24}|^s]
       \right\}_\text{ct}
    = -\nabla^2 R_{a_{1234}} {\cal J}_\phi^{a_{1234}}[|k_{24}|^s] i_0  \cr
   &= (-2{\color[rgb]{0,0.6,0} \nabla_{a_{13}} R_{a_{24}}} + 2{\color{red} \nabla_{a_{12}} R_{a_{34}}})
       {\cal J}_\phi^{a_{1234}}[|k_{24}|^s] i_0 + O(RR) \,,
 }
where the color coding is to show which terms in \eno{FourTwoCounterterms} contribute to which terms of \eno{GradGradSummary}.

Let us briefly summarize the analogous steps for $S_{6,\nabla\nabla R}^{3+3}$.  Obvious relations are
 \eqn{ThreeThreeRelations}{\seqalign{\span\TC}{
  (12) = 0 \qquad (34) = 0 \qquad (13) = (24)  \cr
  (14) = (23) \qquad (15) = (35) = (26) = (46) \qquad
  (16) = (36) = (25) = (45) \,,
 }}
from which it follows that the desired counterterms are
 \eqn{SThreeThreePlan}{
  S_{6,\nabla\nabla R}^{3+3,\text{ct}} &= {\Lambda^\epsilon \over 240\epsilon}
   \left[ 2I_{6,\nabla\nabla R,(13)}^{3+3,\text{ct}}
     + 2 I_{6,\nabla\nabla R,(14)}^{3+3,\text{ct}}
     + 4 I_{6,\nabla\nabla R,(15)}^{3+3,\text{ct}}
     + 4 I_{6,\nabla\nabla R,(25)}^{3+3,\text{ct}}
     + I_{6,\nabla\nabla R,(56)}^{3+3,\text{ct}} \right] \,.
 }
By direct computation using the rules \eno{CountertermRules},
 \eqn{ThreeThreeCounterterms}{
  I_{6,\nabla\nabla R,(13)}^{3+3,\text{ct}} &= {1 \over \tilde\gamma}
    \nabla_{a_{56}} R_{a_{1234}}
     \left\{ {\cal J}^{\contraction{}{a_1}{a_2}{a_3} a_1 a_2 a_3 a_{456}}_\phi[|k_{246}|^s]
       \right\}_\text{ct}
    = -\nabla_{a_{56}} R_{a_{24}} {\cal J}_\phi^{a_{2456}}[|k_{246}|^s] i_0  \cr
   &= -{\color{blue} \nabla_{a_{34}} R_{a_{12}}} {\cal J}_\phi^{a_{1234}}[|k_4|^s] i_0  \cr
  I_{6,\nabla\nabla R,(14)}^{3+3,\text{ct}} &= {1 \over \tilde\gamma}
    \nabla_{a_{56}} R_{a_{1234}} 
     \left\{ {\cal J}^{\contraction{}{a_1}{a_{23}}{a_4} a_1 a_{23} a_4 a_{56}}_\phi[|k_{246}|^s]
       \right\}_\text{ct}
    = \nabla_{a_{56}} R_{a_{23}} {\cal J}_\phi^{a_{2356}}[|k_{26}|^s] i_2  \cr
   &= {\color{red} \nabla_{a_{12}} R_{a_{34}}} {\cal J}_\phi^{a_{1234}}[|k_{24}|^s] i_2  \cr
  I_{6,\nabla\nabla R,(15)}^{3+3,\text{ct}} &= {1 \over \tilde\gamma}
    \nabla_{a_{56}} R_{a_{1234}}
     \left\{ {\cal J}^{\contraction{}{a_1}{a_{234}}{a_5} a_1 a_{234} a_5 a_6}_\phi[|k_{246}|^s]
       \right\}_\text{ct}
    = -\nabla_{ba_6} R^b{}_{a_{234}} {\cal J}_\phi^{a_{2346}}[|k_{246}|^s] i_0  \cr
   &= ({\color{brown} \nabla_{a_{12}} R_{a_{34}}} - {\color{blue} \nabla_{a_{34}} R_{a_{12}}})
      {\cal J}_\phi^{a_{1234}}[|k_4|^s] i_0 + O(RR)  \cr
  I_{6,\nabla\nabla R,(25)}^{3+3,\text{ct}} &= {1 \over \tilde\gamma}
    \nabla_{a_{56}} R_{a_{1234}}
     \left\{ {\cal J}^{a_1 \contraction{}{a_2}{a_{34}}{a_5} a_2 a_{34} a_5 a_6}_\phi[|k_{246}|^s]
       \right\}_\text{ct}
    = -\nabla_{ba_6} R_{a_1}{}^b{}_{a_{34}} {\cal J}_\phi^{a_{1346}}[|k_{46}|^s] i_2  \cr
   &= ({\color{red} \nabla_{a_{12}} R_{a_{34}}} - {\color[rgb]{0,0.6,0} \nabla_{a_{13}} R_{a_{24}}})
      {\cal J}_\phi^{a_{1234}}[|k_{24}|^s] i_2 + O(RR)  \cr
  I_{6,\nabla\nabla R,(56)}^{3+3,\text{ct}} &= {1 \over \tilde\gamma}
    \nabla_{a_{56}} R_{a_{1234}}
     \left\{ {\cal J}^{a_{1234} \contraction{}{a_5}{}{a_6} a_5 a_6}_\phi[|k_{246}|^s]
       \right\}_\text{ct}
    = -\nabla^2 R_{a_{1234}} {\cal J}_\phi^{a_{1234}}[|k_{24}|^s] i_2  \cr
   &= (-2{\color[rgb]{0,0.6,0} \nabla_{a_{13}} R_{a_{24}}} + 2{\color{red} \nabla_{a_{12}} R_{a_{34}}})
      {\cal J}_\phi^{a_{1234}}[|k_{24}|^s] i_2 + O(RR) \,.
 }
Putting \eno{SctPlan}-\eno{FourTwoCounterterms} and \eno{SThreeThreePlan}-\eno{ThreeThreeCounterterms} together and comparing with \eno{GradGradSummary}, we arrive at
 \eqn{KTogether}{
  K_3 &= {4 \over 60} i_0 + {2 \over 60} i_0 + {2 \over 120} i_2 + {4 \over 120} i_2 + 
    {2 \over 120} i_2  \cr
  K_4 &= -{1 \over 60} i_0 - {4 \over 60} i_0 - {2 \over 60} i_0 - {4 \over 120} i_2 -
    {2 \over 120} i_2  \cr
  K_5 &= {2 \over 60} i_2 + {4 \over 60} i_2 + {4 \over 120} i_0  \cr
  K_6 &= -{4 \over 60} i_2 - {2 \over 120} i_0 - {4 \over 120} i_0 \,,
 }
which agrees with \eno{KValues}.

\section{Renormalization at quartic order}
\label{LAPLACIAN}

To renormalize at quartic order, we first inquire whether the counterterms \eno{GradGradSummary} can be organized into the bare arc length action \eno{SBdef}, using the field redefinition \eno{CoordinateMap} and the relationship \eno{gBare} between the bare and renormalized metric.  For the non-local model, we will find that this is impossible!  So we will turn to a generalized form of the bare action that includes a term proportional to the target space laplacian of the square of the arc length.

To get started, we need the bare arc length formula to quartic order in the bare fields: That is, we need one more term in the series \eno{BareArc} than we computed in section~\ref{RENORMALIZATION}.  The computation of arc length is less simple than for the renormalized metric because the $\phi^a_\text{B}$ are not Riemann normal coordinates for $g^\text{B}_{ab}$---due to effects at $O(\tilde\gamma)$, in particular a connection $\Gamma^{\text{B}a}_{b_{12}} \sim O(\tilde\gamma)$.  As a technical device, we therefore introduce a third set of coordinates, $\overline\phi^a$, which are Riemann normal coordinates for the bare metric, which in barred coordinates takes the form $\overline{g}_{ab}(\overline\phi)$.  We can express $\overline\phi^a$ in terms of $\phi_\text{B}^a$ as
 \eqn{BarOverline}{
  \overline\phi^a = \phi_\text{B}^a + \tilde\gamma \left[ L^a{}_{b_{12}} \phi_\text{B}^{b_{12}} + 
    M^a{}_{b_{123}} \phi_\text{B}^{b_{123}} \right] + O(\phi_\text{B}^4) + O(\tilde\gamma^2) \,,
 }
and we can write $g^\text{B}_{ab}(\phi_\text{B})$ in terms of $\overline{g}_{ab}(\overline\phi)$ as
 \eqn{gBarOverline}{
  g^\text{B}_{b_{12}}(\phi_\text{B}) = 
  \overline{g}_{a_{12}}(\overline\phi) 
    {\partial\overline\phi^{a_1} \over \partial\phi_\text{B}^{b_1}}
    {\partial\overline\phi^{a_2} \over \partial\phi_\text{B}^{b_2}} \,.
 }
Note that, by assumption, $g^\text{B}_{a_{12}} = \overline{g}_{a_{12}}$ at $\phi_\text{B} = \overline\phi = 0$.  The condition that $\overline\phi^a$ are Riemann normal coordinates allows us to conclude
 \eqn{LMexpress}{
  \tilde\gamma L^a{}_{b_{12}} = {1 \over 2} \Gamma^{\text{B}a}_{b_{12}} \qquad\qquad
  \tilde\gamma M^a{}_{b_{123}} = {1 \over 6} \left( {\partial \over \partial\phi_\text{B}^{(b_1}} 
    \Gamma^{\text{B}a}_{b_{23})} + \Gamma^{\text{B}a}_{b(b_1} \Gamma^{\text{B}b}_{b_{23})} 
     \right) \,;
 }
see for example \cite{Brewinonline} for a derivation.  The $\Gamma^\text{B} \Gamma^\text{B}$ term in the expression \eno{LMexpress} for $M^a{}_{b_{123}}$ is optional because it is $O(\tilde\gamma^2)$, but it arises naturally in the derivation of \cite{Brewinonline}, so we include it.

The bare arc length coincides between $\overline\phi$ and $\phi_\text{B}$ coordinate systems because these are just different coordinate systems for the same metric, as per \eno{gBarOverline}.  Explicitly,
 \eqn{QBexpand}{
  Q^\text{B}(X_\text{B},Y_\text{B}) 
    = \overline{Q}(\overline{X},\overline{Y})
    = \overline{g}_{a_{12}} (\overline{X}^{a_1} - \overline{Y}^{a_1})
       (\overline{X}^{a_2} - \overline{Y}^{a_2}) - 
       {1 \over 3} \overline{R}_{a_{1234}} \overline{X}^{a_{13}} \overline{Y}^{a_{24}} + 
        O(\overline\phi^5) \,,
 }
where in the second equality we used the fact that $\overline\phi^a$ are Riemann normal coordinates.  The notation $O(\bar\phi^5)$ in \eno{QBexpand} is short for all terms involving five or more powers of $\overline{X}$ and $\overline{Y}$ combined; similar notation is used below.  Using the first equation in \eno{BarOverline} to eliminate $\overline{X}$ and $\overline{Y}$ in favor of $X_\text{B}$ and $Y_\text{B}$, we arrive at
 \eqn{QBexplicit}{
  Q^\text{B}(X_\text{B},&Y_\text{B}) = 
   g^\text{B}_{a_{12}} (X_\text{B}^{a_1} - Y_\text{B}^{a_1}) (X_\text{B}^{a_2} - Y_\text{B}^{a_2})
      \cr &
   + 2 \tilde\gamma \left[ L_{a_{312}} (X_\text{B}^{a_{12}} - Y_\text{B}^{a_{12}})
       (X_\text{B}^{a_3} - Y_\text{B}^{a_3}) 
   + M_{a_{4123}} (X_\text{B}^{a_{123}} - Y_\text{B}^{a_{123}})
      (X_\text{B}^{a_4} - Y_\text{B}^{a_4}) \right]
      \cr &
   - {1 \over 3} R^\text{B}_{a_{1234}} X_\text{B}^{a_{13}} Y_\text{B}^{a_{24}}
   + O(\phi_\text{B}^5) + O(\tilde\gamma^2) \,,
 }
where $L_{a_{312}} = g^\text{B}_{a_3b} L^b{}_{a_{12}}$ and $M_{a_{4123}} = g^\text{B}_{a_4b} M^b{}_{a_{123}}$.  Note that the cubic terms in \eno{QBexplicit} agree with \eno{CubicQ}, and recall from the subsequent analysis that the corresponding cubic term $S^\text{B}_3$ does {\it not} need to match $S_3 + S^\text{ct}_3$, because of the additional term cubic in $\phi$ in \eno{StwoExpand} arising from the $O(\tilde\gamma)$ difference between $\phi_\text{B}$ and $\phi$.
Likewise, the term in \eno{QBexplicit} proportional to $M_{a_{4123}}$ gives rise to a term proportional to ${\cal J}_{\phi_\text{B}}[|k_4|^s]$ in the action, but we do not need to track it explicitly because the quartic term in \eno{StwoExpand} shows that it is precisely the sort of term that we can absorb into the field redefinition coefficient $W^a{}_{b_{123}}$.  Thus we may write
 \eqn{SFourBare}{
  S^\text{B}_4 &= {\Lambda^\epsilon \over 2\gamma} \int' {d^n x d^n y \over |x-y|^{n+s}}
    Q^\text{B}_4(\phi_\text{B}(x),\phi_\text{B}(y)) = 
    {\Lambda^\epsilon \over 12 \hat\gamma} 
    {\cal J}_{\phi_\text{B}}^{a_{1234}}[|k_{24}|^s] R^\text{B}_{a_{1234}} + \text{(field redef)} \,,
 }
where $\text{(field redef)}$ indicates field redefinition terms as discussed above.

Next we express $S^\text{B}_4$ in terms of renormalized quantities in order to compare to \eno{GradGradSummary}.  Starting from \eno{gBare}, we obtain
 \eqn{RiemannRelation}{
  \left( {\Lambda \over \mu} \right)^\epsilon R^\text{B}_{a_{1234}}
    &= R_{a_{1234}} + {\tilde\gamma \over 2} (
      \nabla_{a_{14}} T_{a_{23}} - \nabla_{a_{24}} T_{a_{13}}
    - \nabla_{a_{13}} T_{a_{24}} + \nabla_{a_{23}} T_{a_{14}})
   \cr &\qquad\qquad{}
     + O(RR) + O(\tilde\gamma^2) \,.
 }
Thus we find
 \eqn{SFourBareAgain}{
  S^\text{B}_4 &= S_4 + {\Lambda^\epsilon \over 12\epsilon}
    {\cal J}_{\phi}^{a_{1234}}[|k_{24}|^s] 
     \left[ t_0 {\color{red} \nabla_{a_{12}} R_{a_{34}}} - 
      t_0 {\color[rgb]{0,0.6,0} \nabla_{a_{13}} R_{a_{24}}} \right]
   \cr &\qquad\qquad{}
   + \text{(field redef)} + O(RR) + O(\tilde\gamma) \,.
 }
$O(RR)$ terms arise in \eno{SFourBareAgain} not just from those in \eno{RiemannRelation}, but also from expressing ${\cal J}_{\phi_\text{B}}^{a_{1234}}[|k_{24}|^s]$ in terms of the renormalized field $\phi$.\footnote{$S^\text{B}_3$ does not contribute terms quartic in $\phi$ that we need to track because it starts at $O(\tilde\gamma^0)$, so quartic terms coming from expressing ${\cal J}_{\phi_\text{B}}^{a_{123}}[|k_3|^s]$ in terms of the renormalized $\phi$ enter at $O(\tilde\gamma)$.}  We have color-coded terms in \eno{SFourBareAgain} to match the way we did in \eno{GradGradSummary}.  Comparing the two equations, we can see that $S^\text{B}_4$ accommodates the counterterms $S^\text{ct}_{6,\nabla\nabla R}$ iff $K_3 = -K_4$.  Based on \eno{KValues}, this happens iff $i_0=i_2$, which means iff $s$ is a positive even integer.  As we will develop more fully in section~\ref{LOCAL}, this indeed corresponds to the case of local non-linear sigma models. 

Let us pursue further here what happens for the non-local case.  Because $i_2=0$,  we have $K_1/K_2 = -6/7$, there is no hope of rendering the theory renormalizable with just the arc length action we have been using so far.  Some generalization of the arc length action is needed.  Whatever modification we make should involve two target space derivatives relative to the original action, so as to absorb counterterms that appear with two extra derivatives---like the $\nabla_{a_{12}} R_{a_{34}}$ structure in $S^\text{ct}_{6,\nabla\nabla R}$ as compared to $R_{a_{1234}}$ in $S_4$.  Luckily, there is a new term with the right properties which we can add to $S^\text{B}$:
 \eqn{deltaSpp}{
  \delta S^\text{B} = \kappa_\text{B} {\Lambda^\epsilon \over 2\gamma} 
    \int' {d^n x d^n y \over |x-y|^{n+s}} 
      Q^{\text{B}\prime\prime}(\phi_\text{B}(x),\phi_\text{B}(y))
 }
where we define
 \eqn{QBpp}{
  Q^{\text{B}\prime\prime}(X_\text{B},Y_\text{B}) \equiv 
    (\nabla_{X_\text{B}}^2 + \nabla_{Y_\text{B}}^2) Q^\text{B}(X_\text{B},Y_\text{B}) - 
      4n \,.
 }
The $-4n$ term in \eno{QBpp} is present in order to ensure $Q^{\text{B}\prime\prime}(X_\text{B},Y_\text{B}) = 0$ when $X_\text{B}$ and $Y_\text{B}$ coincide.\footnote{Strictly speaking we do not need the $-4n$ subtraction in \eno{QBpp} when we used a regulated integral in \eno{deltaSpp}, so including it explicitly in \eno{QBpp} is a matter of taste.}  By explicit calculation (as sketched below \eno{QppsixCompute}),
 \eqn{deltaSppCompute}{
  \delta S^\text{B} &= \kappa_\text{B} {\Lambda^\epsilon \over 60\gamma} 
    (-6 \nabla^\text{B}_{a_{12}} R^\text{B}_{a_{34}} 
     + 7 \nabla^\text{B}_{a_{13}} R^\text{B}_{a_{24}})
   \cr &\qquad\qquad{} \times
    \int' {d^n x d^n y \over |x-y|^{n+s}} 
      \left[ \phi_\text{B}^{a_{13}}(x) - \phi_\text{B}^{a_{13}}(y) \right]
      \left[ \phi_\text{B}^{a_{24}}(x) - \phi_\text{B}^{a_{24}}(y) \right] 
   \cr &\qquad{}
    + \text{(field redef)} + O(\phi_\text{B}^5) + O(RR) \,,
 }
Passing to momentum space, we find
  \eqn{SppPlugged}{
   \delta S^\text{B} &= \kappa_\text{B} {\Lambda^\epsilon \over 60\hat\gamma}
    (-6 \nabla^\text{B}_{a_{12}} R^\text{B}_{a_{34}} 
     + 7 \nabla^\text{B}_{a_{13}} R^\text{B}_{a_{24}})
    {\cal J}_{\phi_\text{B}}^{a_{1234}}[|k_{24}|^s]
   \cr &\qquad{}
    + \text{(field redef)} + O(\phi_\text{B}^5) + O(RR) \,.
  }
Combining \eno{SFourBareAgain} and \eno{SppPlugged} and comparing with \eno{GradGradSummary}, we see that $\kappa_\text{B} = -{\hat\gamma \over \epsilon} {i_0 \over 2}$.  Strikingly, we are forced also to choose $t_0=0$: That is, the metric is not renormalized!

Having allowed the two-derivative term \eno{deltaSpp} in $S^\text{B}$, we should allow addition of a similar term to the renormalized action:
 \eqn{QppRen}{
  \delta S = \kappa {\mu^\epsilon \over 2\gamma}
    \int' {d^n x d^n y \over |x-y|^{n+s}} Q''(\phi(x),\phi(y)) \,.
 }
We restrict $\kappa$ to be an $O(\gamma)$ quantity, which makes sense because then the overall scaling with $\gamma$ of \eno{QppRen} is $O(\gamma^0)$, and this aligns with the invariance of $Q''(X,Y)$ under overall rescaling of the target manifold.  The additional term \eno{QppRen} changes all the one-loop amplitudes, but only by $O(\gamma)$ quantities, relative to the $O(\gamma^0)$ scaling of one-loop amplitudes and counterterms that we obtained in previous sections.  In short, the only effect of allowing non-zero $\kappa$ in our counterterm analysis is to lead to a direct tree-level contribution to $\kappa_\text{B}$, so that in total,
 \eqn{kappaRelation}{
  \left( {\Lambda \over \mu} \right)^\epsilon \kappa_\text{B} = \kappa 
    - \tilde\gamma {i_0 \over 2} \,.
 }
To rephrase this result in terms of the renormalization group, we can rewrite \eno{kappaRelation} as
 \eqn{kRrewrite}{
  \Lambda^\epsilon \left( \kappa_\text{B} + {\hat\gamma \over \epsilon}
    {i_0 \over 2} \right) = \mu^\epsilon \kappa \,,
 }
and then since the right hand side is independent of $\Lambda$, we arrive at
 \eqn{KappaRuns}{
  \Lambda {d\kappa_\text{B} \over d\Lambda} = -\epsilon \kappa_\text{B} - 
   \hat\gamma {i_0 \over 2} \,.
 }
The first term on the right hand side of \eno{KappaRuns} is the tree-level term coming from the engineering dimension factor of $(\Lambda/\mu)^\epsilon$ in \eno{kappaRelation}.  The one-loop effects are responsible for the second term in \eno{KappaRuns}.  If we now take $\epsilon \to 0$ in \eno{KappaRuns}, we see that $\kappa_\text{B}$ runs logarithmically:
 \eqn{KappaLog}{
  \kappa_\text{B} = -\hat\gamma {i_0 \over 2} \log {\Lambda \over \Lambda_0} \,,
 }
where $\Lambda_0$ is a dynamically generated scale.  Note that $\hat\gamma$ and $i_0$ are positive, so $\kappa_\text{B}$ is positive at scales $\Lambda$ below $\Lambda_0$ and negative above $\Lambda_0$.

To see that \eno{deltaSppCompute} is correct, let's work on the renormalized side and note that
 \eqn{QppsixCompute}{
  g^{b_{12}} {\partial^2 \over \partial X^{b_1} \partial X^{b_2}} Q_6^{\nabla\nabla R} &= 
    -{1 \over 60} \nabla_{a_{56}} R_{a_{1234}} Y^{a_{24}}
     g^{b_{12}} {\partial^2 X^{a_{1356}} \over \partial X^{b_1} \partial X^{b_2}} + \dots  \cr
   &= -{1 \over 30} (\nabla_{a_{13}} R_{a_{24}} + 4 \nabla_{a_1b} R^b{}_{a_{234}} + 
     \nabla^2 R_{a_{1234}}) X^{a_{13}} Y^{a_{24}} + \dots  \cr
   &= -{1 \over 30} (6 \nabla_{a_{13}} R_{a_{24}} - 6 \nabla_{a_{12}} R_{a_{34}} + 
     \nabla_{a_{24}} R_{a_{13}}) X^{a_{13}} Y^{a_{24}} + \ldots \,.
 }
The expression \eno{QppsixCompute} is part of $\nabla_X^2 Q(X,Y)$, and it is easy to see that it is the only part contributing terms of the form $(\nabla\nabla R) XXYY$.  Therefore
 \eqn{QppExplicit}{
  Q''(X,Y) &= -{1 \over 30} (-6 \nabla_{a_{12}} R_{a_{34}} + 7 \nabla_{a_{13}} R_{a_{24}}) 
    (X^{a_{13}} Y^{a_{24}} + X^{a_{24}} Y^{a_{13}}) 
   \cr &\qquad{} + O(\phi^5) + O(RR) + \ldots \,.
 }
The ellipses in \eno{QppsixCompute} and \eno{QppExplicit} indicate terms that are not quadratic in both $X$ and $Y$, for example terms schematically of the form $(\nabla\nabla R) XYYY$ or $(\nabla\nabla R) YYYY$, as well as lower order terms which are either independent of $X$ or $Y$, or linear in $X$ or $Y$.  Plugging \eno{QppExplicit} into \eno{QppRen}, we find
 \eqn{deltaSppCross}{
  \delta S &= -\kappa {\mu^\epsilon \over 60\gamma} (-6 \nabla_{a_{12}} R_{a_{34}} + 
    7 \nabla_{a_{13}} R_{a_{24}})  \cr
   &\qquad\qquad{} \times \int' {d^n x d^n y \over |x-y|^{n+s}}
    \left[ \phi^{a_{13}}(x) \phi^{a_{24}}(y) + \phi^{a_{24}}(x) \phi^{a_{13}}(y) \right]
    \cr &\qquad{} + \text{(field redef)} + O(\phi^5) + O(RR)  \cr
   &= \kappa {\mu^\epsilon \over 60\gamma} (-6 \nabla_{a_{12}} R_{a_{34}} + 
    7 \nabla_{a_{13}} R_{a_{24}})  \cr
   &\qquad\qquad{} \times \int' {d^n x d^n y \over |x-y|^{n+s}}
    \left[ \phi^{a_{13}}(x) - \phi^{a_{13}}(y) \right]
    \left[ \phi^{a_{24}}(x) - \phi^{a_{24}}(y) \right]
    \cr &\qquad{} + \text{(field redef)} + O(\phi^5) + O(RR) \,.
 }
The contributions labeled $\text{(field redef)}$ in \eno{deltaSppCross} are linear in $\phi(x)$ or $\phi(y$).  To see that \eno{deltaSppCross} is correct, we have only to understand why we can freely add or drop from the integrand smooth functions which depend only on $x$ or only on $y$, such as the direct terms $\phi^{a_{1234}}(x) + \phi^{a_{1234}}(y)$ which are present in the last expression in \eno{deltaSppCross} but not the middle expression.  As in section~\ref{BILOCAL}, this follows from careful use of the regulated integral prescription:
 \eqn{VanishingInt}{
  \int' {d^n x d^n y \over |x-y|^{n+s}} f(y) &= 
    \int {d^n x d^n y \over |x-y|^{n+s}} \Bigg[ f(y)
    - \sum_{r=0}^{\lfloor s/2 \rfloor}
       b_r \square^r f(x) (y-x)^{2r}  \Bigg]  \cr
    &= {1 \over \Gamma_V(n+s)} \int d^n x \, D^s f(x) = 0
 }
for smooth functions $f(x)$ with suitable falloff conditions at large $x$.  The integral $\int' {d^n x d^n y \over |x-y|^{n+s}} f(x)$ vanishes more trivially by subtraction of the $r=0$ term in the sum appearing in square brackets in \eno{VanishingInt}.

\section{Recovering the local non-linear sigma model}
\label{LOCAL}

When $V = \mathbb{R}^n$ and $s$ is a positive even integer, the original position space action \eno{Sdef} is problematic, because our scheme for defining regulated integrals breaks down.  However, this is precisely the case where we expect to recover a local theory, expressible in terms of $\phi(x)$ and its derivatives.  In this section we will verify this expectation for the simplest case, $s=2$.  We proceed in three steps: First, we start from the momentum space form of the action, obtained for general $s$ from the non-local action \eno{Sdef}; then we take $s=2$; and finally we Fourier transform back to a local position-space action.  It is helpful (though not really necessary) to work with Riemann normal coordinates throughout.  We exhibit the process in detail for the quadratic and quartic terms in the action.  The first step is to put \eno{SfreeAgain} and \eno{SquarticFourier} together to get
 \eqn{STwoFour}{
  S_2 + S_4 &= {\mu^\epsilon \over 2\hat\gamma} g_{a_{12}} 
      \int d^n k \, \hat\phi^{a_1}(-k) |k|^s \hat\phi^{a_2}(k)
    \cr &\qquad{}
    + {\mu^\epsilon \over 12\hat\gamma} R_{a_{1234}} \int d^n k \, 
    (\widehat{\phi^{a_1} \phi^{a_3}})(-k) |k|^s (\widehat{\phi^{a_2} \phi^{a_4}})(k) \,.
 }
Having set $s=2$, we pass to position space, keeping in mind that in our conventions, $k^2 \to -{1 \over 4\pi^2} \square$.  The result is
 \eqn{StfPosition}{
  S_2 + S_4 &= {\mu^\epsilon \over 8\pi^2 \hat\gamma} \int d^n x 
    \Bigg[ g_{a_{12}} \partial_{x^i} \phi^{a_1}(x) \partial^{x^i} \phi^{a_2}(x)
    \cr &\qquad\qquad{}
     + {1 \over 6} R_{a_{1234}} \partial_{x^i} (\phi^{a_1}(x) \phi^{a_3}(x))
       \partial^{x^i} (\phi^{a_2}(x) \phi^{a_4}(x)) \Bigg] \,,
 }
where integration by parts was allowed without generating derivatives of $g_{a_{12}}$ or $R_{a_{1234}}$ because these tensors are evaluated at $\phi=0$.  The result \eno{StfPosition} is to be compared with the standard local sigma model action, expanded to quartic order in fields:
 \eqn{SNLsM}{
  S_\text{local} &= {\mu^\epsilon \over 2\sigma} \int d^n x \, g_{a_{12}}(\phi(x))
    \partial_{x^i} \phi^{a_1}(x) \partial^{x^i} \phi^{a_2}(x)
   \cr &
   = {\mu^\epsilon \over 2\sigma} \int d^n x \left[ g_{a_{12}} -
     {1 \over 3} R_{a_{1324}} \phi^{a_3}(x) \phi^{a_4}(x) + O(\phi^3) \right] 
    \partial_{x^i} \phi^{a_1}(x) \partial^{x^i} \phi^{a_2}(x) \,.
 }
It is easy to check that \eno{StfPosition} and \eno{SNLsM} agree provided $\sigma = 4\pi^2 \hat\gamma$.  Note that $\gamma = -{2 \over \Gamma_{\mathbb{R}^n}(n+2)} \hat\gamma$ has a simple pole at $s=2$.  This is a reminder that the original action \eno{Skin} is ill-defined when $s=2$.  However, the $s \to 2$ limit is well-defined and smooth in Fourier space, with $\hat\gamma$ held fixed and positive.

It is straightforward but unilluminating to extend the comparison of \eno{StfPosition} and \eno{SNLsM} through $O(\phi^6)$; suffice it to say that one does recover $S_\text{local}$ order by order in $\phi$ through the procedure outlined in the previous paragraph.

Let's now use the results of previous sections to recover the well-known beta function for the local non-linear sigma model on $V = \mathbb{R}^n$ with $s=2$ and $n=2+\epsilon$, for $\epsilon$ positive and sufficiently small.  Comparing the first line of \eno{GradGradSummary} with the first line of \eno{SFourBareAgain}, we conclude that the metric renormalization coefficient $t_0$ must take the value
 \eqn{tZeroValue}{
  t_0 = 6K_3 = i_0 = {2 \over \zeta_\infty(n)} = 2\pi \,,
 }
where we used $i_2=i_0$ from \eno{iTwoCompute} together with \eno{KValues} to conclude that $K_3 = -K_4 = i_0/6$; then in the third equality of \eno{tZeroValue} we used \eno{IzeroAgain} to obtain the explicit value of $i_0$ in dimension $n=2+\epsilon$.  In \eno{tZeroValue}, we are only tracking $i_0$ and $i_2$ up to $O(\epsilon)$ contributions, consistent with our aim of calculating only the leading behavior in $\epsilon$ of the one-loop beta function.  It is worth noting that in \eno{tZeroValue}, we are extracting $t_0$ exclusively from the four-point amplitude.  The two-point amplitude (i.e.~the one-loop correction to the propagator) doesn't help at all in determining $t_0$; instead, it results in the condition \eno{stiConstraint} which determines the field renormalization coefficient $u_0$ once $t_0$ is known---specifically, $u_0 = -i_0/3$.  An important point is that we are able to completely cancel one-loop divergences with metric renormalization (together with field redefinition), consistent with the local non-linear sigma model being renormalizable. 

To go further and extract the beta function for the metric, we start with a rearrangement of \eno{gBare}:
 \eqn{gBareAgain}{
  \Lambda^\epsilon \left[ g_{ab}^\text{B}(\phi_\text{B}) - 
    {\hat\gamma \over \epsilon} t_0 R^\text{B}_{ab}(\phi_\text{B}) \right] = 
    \mu^\epsilon g_{ab}(\phi) + O(\tilde\gamma^2) \,.
 }
Let's set $\phi = \phi_\text{B} = 0$ and differentiate both sides with respect to $\Lambda$, holding renormalized quantities fixed.  The result is
 \eqn{gabRG}{
  \Lambda {dg^\text{B}_{ab} \over d\Lambda} = -\epsilon g^\text{B}_{ab} + 
    \hat\gamma t_0 R^\text{B}_{ab} \,.
 }
Using $\hat\gamma = \sigma/4\pi^2$ and $i_0 = 2\pi$, we arrive at
 \eqn{gabSpecific}{
  \Lambda {dg^\text{B}_{ab} \over d\Lambda} = -\epsilon g^\text{B}_{ab} + 
    {\sigma \over 2\pi} R^\text{B}_{ab} \,.
 }
This accords with the standard one-loop result as quoted for example in \cite{Friedan:1980jm}.\footnote{When using the results of \cite{Friedan:1980jm}, one must keep in mind that the parameter $T$ (analogous to our $\sigma$) appearing in the beta function $\beta_{ij}(T^{-1} g) = -\epsilon T^{-1} g_{ij} + R_{ij}$ must be replaced by $T/2\pi$, so that $\beta_{ij}(g) = -\epsilon g_{ij} + {T \over 2\pi} R_{ij}$, in order to be consistent with the action $S = \Lambda^\epsilon \int dx \, {1 \over 2} T^{-1} g_{ij}(\phi(x)) \partial_\mu \phi^i(x) \partial_\mu \phi^j(x)$; see the comment in this regard on p.~388.}

For positive even $s>2$, the action of a local, $s$-derivative non-linear sigma model does not appear to be uniquely determined by symmetries (though it is possible we do not fully understand how diffeomorphism symmetry is implemented in these higher derivative theories).  The simplicity of our loop calculations for all positive even $s$ suggests that there may be a privileged local non-linear sigma model in each positive even dimension whose renormalization group flow is characterized by \eno{gabRG}.

\section{Outlook}
\label{OUTLOOK}

Let's start with a recapitulation of the main points of our analysis.  The starting point action is
 \eqn{Srecap}{
  S = {\mu^\epsilon \over 2\gamma} \int'_{xy} Q(\phi(x),\phi(y)) \,,
 }
where $Q(X,Y) = d(X,Y)^2$ is the square of the shortest distance between points $X$ and $Y$ on the target manifold, and we understand that
 \eqn{ShortInt}{
  \int'_{xy} G(x,y) = \int'_{V \times V} {d^n x d^n y \over |x-y|^{n+s}} G(x,y)
 }
is defined with a suitable regulation prescription, as in section~\ref{BILOCAL}.  Recall that $\epsilon=n-s$.  Focusing on the limit $\epsilon \to 0^+$, with $n$ and $s$ converging to some positive $n_0$ which is not an even integer when $V = \mathbb{R}^n$, we find that we are obliged to generalize the action \eno{Srecap} to
 \eqn{Simproved}{
  S_\text{improved} = {\mu^\epsilon \over 2\gamma} \int'_{xy} \left[ Q(\phi(x),\phi(y)) + 
    \kappa Q''(\phi(x),\phi(y)) \right] \,,
 }
where $Q''(X,Y) = (\nabla_X^2 + \nabla_Y^2) Q(X,Y)$, and $\kappa \sim O(\gamma)$.  With this improved action, one-loop amplitudes at $O(\gamma^0)$ have a divergence structure which, as far as we have taken the computations, can be absorbed entirely through field redefinitions and additive renormalization of $\kappa$, as given in the form of a renormalization group equation for the bare version of $\kappa$ in \eno{KappaRuns}.  No metric renormalization arises in the non-local model (at one loop).  This is in contrast with the local non-linear sigma model, where no improvement terms are needed, and we cancel one-loop divergences instead through field redefinitions and renormalization of the metric.

The one-loop divergences we encountered are all proportional to the Ricci tensor---more precisely, to covariant derivatives of it.  So all of them vanish when $R_{ab}=0$.  However, we did not analyze diagrams proportional to the square of the Riemann tensor.  In the local case, the sum of the divergences of all diagrams (at one loop) vanishes when $R_{ab}=0$.  It would seem sensible to find the same outcome for the non-local case, but settling this question is left for future work.

In the non-local theory \eno{Srecap}, we expect that additional improvement terms will be needed at each new order in $\gamma$, corresponding to higher order target space derivatives of $Q(X,Y)$.  This would be somewhat analogous to the situation in chiral perturbation theory: At each order in a derivative expansion one has finitely many parameters to adjust, but every new order introduces new free parameters.  However, an alternative viewpoint may be possible: Instead of renormalizing individual parameters like $\kappa$, it may be possible to renormalize $Q(X,Y)$ as a bi-local function, regarding the improvement term $\kappa Q''(X,Y)$ as an additive renormalization of $Q(X,Y)$.  To find a beta function for $Q$ even at the one-loop level, we would need full control over all field redefinitions, because $\Lambda {d \over d\Lambda} Q^\text{B}(X_\text{B},Y_\text{B})$ should be computed with the renormalized $X$ and $Y$ held fixed, and that means $X_\text{B}$ and $Y_\text{B}$ will have some $\Lambda$ dependence.  If this alternative viewpoint is indeed viable, then the relation $Q(X,Y) = d(X,Y)^2$ would be only an initial condition that we might impose at some scale.  Some technical aspects of our analysis might have to change to accommodate a function $Q(X,Y)$ that is not locked to the square of the distance function; in particular, $Q_3(X,Y)$ may not vanish in any convenient coordinate system.

\begin{figure}
  \centerline{\includegraphics[width=2.5in]{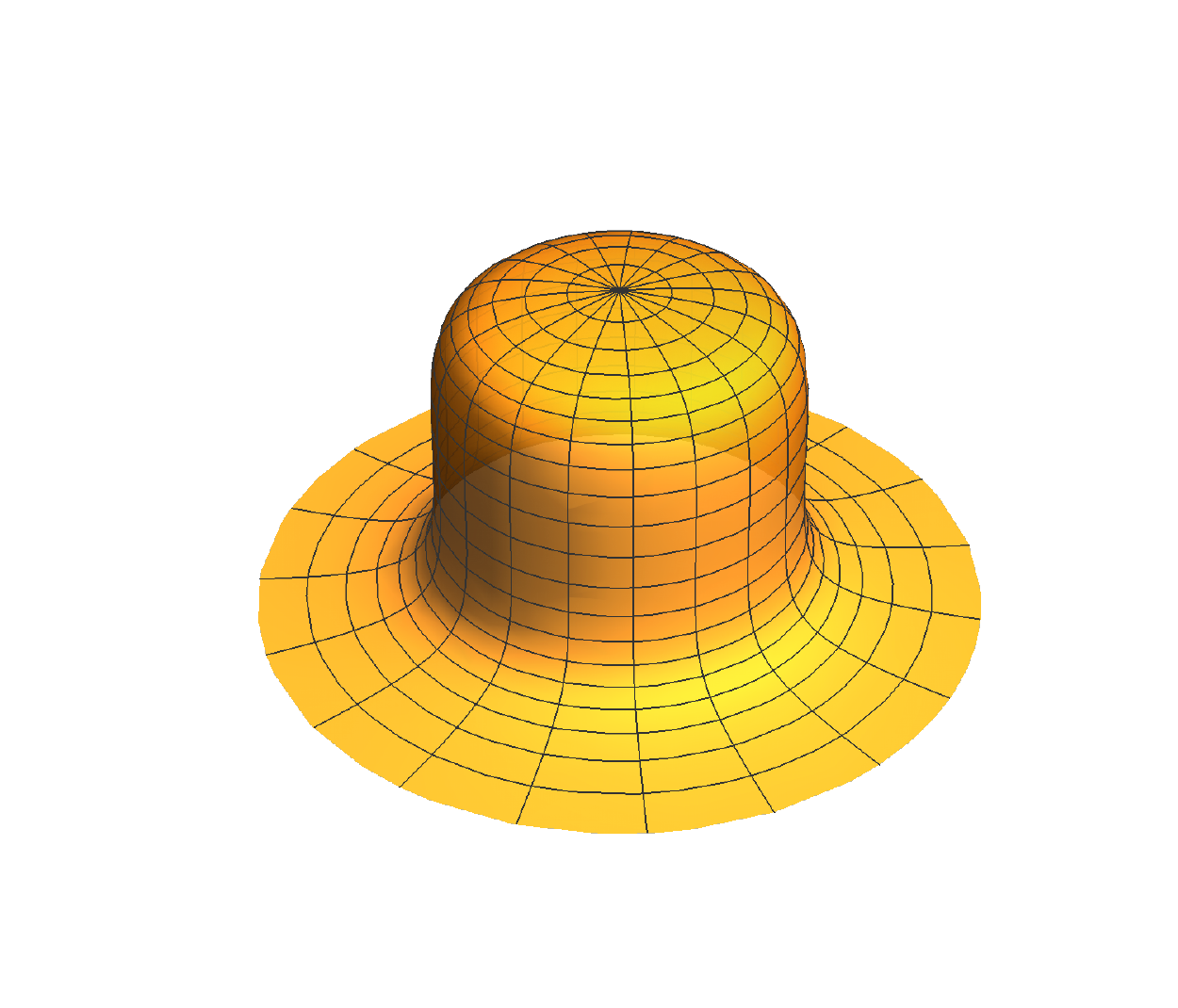}}
  \caption{A surface with a bulge which is cylindrical over some sufficient distance and then closes off.  The distance function $d(X,Y)^2$ on this surface will have discontinuities in its first derivative.}\label{RaggedExamples}
 \end{figure}

There is a reason already at the level of the classical action to consider relaxing the relation $Q(X,Y) = d(X,Y)^2$.  Namely, $d(X,Y)^2$ is in general not a globally smooth function, though it is certainly smooth when $X$ and $Y$ are sufficiently close to one another.  Consider for example making the target space manifold $M$ a circle parametrized by an angle $\theta \in \mathbb{R} \mod 2\pi$.  Then $d(0,\theta)^2 = \theta^2$ for $\theta \in [-\pi,\pi]$, but right at $\pi$ one finds a discontinuity in the first derivative of $d(0,\theta)^2$, so that e.g.~$d(0,\theta) = (\theta-2\pi)^2$ for $\theta \in [\pi,3\pi]$.  This example makes it seem like failure of differentiability of $Q(X,Y)$ is associated with a homology cycle, but that is not necessarily the case.  For example, consider first a cylinder, $M = S^1 \times \mathbb{R}$.  Clearly, $d(X,Y)^2$ again has a discontinuity in its first derivative when $X$ and $Y$ are diametrically opposite points.  Now consider an asymptotically flat surface which has a sufficiently long cylindrical region so that the same discontinuity in the first derivative of $d(X,Y)^2$ shows up for diametrically opposite points in the cylindrical region.  The upshot is that if an asymptotically flat, smooth, contractible surface has a sufficiently prominent bulge, $d(X,Y)^2$ will not be a smooth function. See figure~\ref{RaggedExamples}.  But---we speculate---renormalization group flow of $Q(X,Y)$ might smooth away the raggedness of $d(X,Y)^2$ without entirely losing the notion of a useful distance function between points on $M$.

There are some obvious extensions of our work which do not depend on renormalizing $Q(X,Y)$ as a function.  First, it would be desirable to work out all the $O(RR)$ terms in one-loop renormalization through four-point amplitudes.  From a diagrammatic point of view, it seems natural that $(\nabla_X^2 + \nabla_Y^2) Q(X,Y)$ should compensate for divergences arising from tying two legs of a vertex with a single propagator.  This is because when the derivatives act as $\partial_X^2$ or $\partial_Y^2$, they eliminate two factors of $X$ from a vertex operator, or two factors of $Y$, just like a Wick contraction.  But this intuition leaves out the covariance of $\nabla_X$ and $\nabla_Y$, and it doesn't explain how to compensate for divergences from the type of diagram shown in figure~\ref{QuarticDiagrams}c.  We hope that some improved understanding is possible that will make it obvious that the detailed calculations in section~\ref{FOUR} had to lead to counterterms that can be assembled into $Q''(X,Y)$.  Perhaps then it will be easier to track field redefinition terms, $O(RR)$ terms, and even higher loop counterterms.

Another obvious extension is to consider more carefully the local limit.  As commented in section~\ref{LOCAL}, we have made no attempt to work out local theories with more than two derivatives, though we note that at least the four-derivative $O(N)$ model in $n=4$ dimensions is a long-established theory \cite{Gava:1978gd}.  We restricted attention to a particular way of taking the $\epsilon \to 0^+$ limit for local theories, namely by setting $s$ exactly equal to a positive even integer and then sliding $n = s+\epsilon$ down toward $s$.  Also, we have not considered possible competition between local and non local theories. For example, if $s<2$ then a local, $s=2$, action is classically irrelevant and one would expect that it is not generated along the renormalization group flow. Borrowing from the analysis of \cite{Fisher:1972zz,PhysRevB.8.281,Honkonen_1989} (see also, e.g., \cite{Paulos:2015jfa,Behan:2017emf} for a modern perspective) it is possible that non local theories with $s<2$ will flow to local theories in the infrared due to quantum corrections.

Finally, we hope that there may be some applicability of non-local theories to a fundamental issue in M-theory, namely the description of M2-branes far from the limit of nearly flat brane configurations.  The M2-brane appears to be afflicted by a pathology which is ultraviolet with respect to the brane worldvolume but infrared from the point of view of spacetime \cite{deWit:1988wri,deWit:1988xki}: It can throw out long thin tendrils with essentially zero action.  We speculate that this issue might find its cure through a modification of the M2-brane in the ultraviolet into a non-local action, perhaps some supersymmetrization of
 \eqn{BraneAction}{
  S &= \int' d^3 x \sqrt{\det g_\text{M2}(x)} \, d^3 y \sqrt{\det g_\text{M2}(y)} \,
    {d_\text{11d}(\phi(x),\phi(y))^2 \over d_\text{M2}(x,y)^6}  \cr
    &\sim \int' {d^3 x d^3 y \over |x-y|^6} d_\text{11d}(\phi(x),\phi(y))^2 \,,
 }
where $d_\text{M2}$ is distance with respect to the intrinsic world-volume metric and $d_\text{11d}$ is distance with respect to the spacetime metric.  The second line of \eno{BraneAction} shows an approximate gauge-fixed form of the action for an M2-brane that is stretched out nearly flat in a large eleven-dimensional geometry.  The power $|x-y|^6$ corresponds to $\epsilon=0$, i.e.~to the value required for the theory to be conformal if loop divergences are absent.  It is our hope that some supersymmetric version of our calculations can be carried out, starting from an action similar to \eno{BraneAction}, to recover eleven-dimensional supergravity as a condition of conformal invariance.  A theory like \eno{BraneAction} is at best a UV description of the M2-brane, and one should anticipate the addition of a two-derivative term, which in gauge-fixed form would read
 \eqn{RelevantBrane}{
  \delta S = {\tau_\text{M2} \over 2} 
    \int d^3 x \, g^\text{11d}_{a_{12}}(\phi(x)) \partial_{x^i} \phi^{a_1}(x) 
    \partial^{x^i} \phi^{a_2}(x) \,.
 }
This term is relevant by the power counting based on \eno{BraneAction}, allowing us to recover our ordinary understanding of the M2-brane at long distances compared to the eleven-dimensional Planck scale.

A theory on the M2-brane world-volume should be capable of being framed in Lorentzian signature.  At least naively, one can make the replacement
 \eqn{xyReplace}{
  |x-y|^6 \to \left[ -(x^0-y^0)^2 + (\vec{x}-\vec{y})^2 + i\varepsilon \right]^3
 }
in \eno{BraneAction}, where $\varepsilon$ is a small positive number (unrelated to $\epsilon = n-s$).  The good news is that the right hand side of \eno{xyReplace} is essentially real---as contrasted for example with $\left[ -(x^0-y^0)^2 + (\vec{x}-\vec{y})^2 + i\varepsilon \right]^{n+s}$ for general $n$ and $s$, which has a phase $e^{i(\pi-\varepsilon)(n+s)}$ inside the lightcone.  Questions abound which we have not explored.  Can causality be maintained?  Is there a relationship with generalized free field theory when the target space geometry is flat?  How does one handle gauge-fixing and the related ghosts?  Is there a canonical formulation of Lorentzian bi-local theories?

\section*{Acknowledgments}

This work was supported in part by the Department of Energy under Grant No.~DE-FG02-91ER40671, and by the Simons Foundation, Grant 511167 (SSG).  In addition, AY is supported in part by an Israeli Science Foundation excellence center grant 2289/18 and a Binational Science Foundation grant 2016324.  We thank C.~Callan, D.~Gross, M.~Heydeman, and I.~Klebanov for useful discussions.

\clearpage

\bibliographystyle{utphys}
\bibliography{NLNLsM}

\end{document}